\documentstyle[12pt]{article}
\catcode`\@=11
\def\numberbysection{\@addtoreset{equation}{section}
        \def\theequation{\thesection.\arabic{equation}}}
\numberbysection

\begin{document}

\newlength{\lno} \lno1.5cm \newlength{\len} \len=\textwidth%
\addtolength{\len}{-\lno}

\setcounter{page}{0} 
\begin{titlepage} 
\vspace{0.5cm}
\begin{center}
{\Large\bf Bethe Ansatz solutions for Temperley-Lieb Quantum Spin Chains }\\
\vspace{1cm}
{\large R. C. T.  Ghiotto and A. L. Malvezzi }\\ \vspace{1cm}
{\large \it Universidade Estadual Paulista - UNESP Faculdade de Ci\^encias Departamento de F\'{\i}sica 
Caixa Postal 473, CEP 17033-360 ~~ Bauru , SP, Brazil}\\
\end{center}
\vspace{1.2cm}

\begin{abstract}
We solve the spectrum of quantum spin chains based on representations of the
Temperley-Lieb algebra associated with the quantum groups ${\cal U}%
_{q}(X_{n})$ for $X_{n}=A_{1},$ $B_{n},$ $C_{n}$ and $D_{n}$. The tool is a
modified version of the coordinate Bethe Ansatz through a suitable choice of
the Bethe states which give to all models the same status relative to their
diagonalization. All these models have equivalent spectra up to degeneracies
and the spectra of the lower dimensional representations are contained in
the higher-dimensional ones. Periodic boundary conditions, free boundary
conditions and closed non-local boundary conditions are considered. Periodic
boundary conditions, unlike free boundary conditions, break quantum group
invariance. For closed non-local cases the models are quantum group
invariant as well as periodic in a certain sense.
\end{abstract}
\vfill{}
\end{titlepage}
\newpage

\renewcommand{\thefootnote}{\arabic{footnote}} \setcounter{footnote}{0}

\section{\sf Introduction}

{\sf Temperley-Lieb (TL) algebra \cite{TL} has been widely used in the
construction of solutions of the Yang-Baxter equation \cite{B, BK}, which is
a sufficient condition for integrability of lattice models through the
quantum inverse scattering method. On the other hand, quantum group\cite{PS}
in some cases is a symmetry of the integrable model \cite{BNMR}. Thus,
quantum group together with Temperley-Lieb algebra provide a powerful
algebraic framework to build and study various kinds of lattice models in
two-dimensional statistical mechanics. One particular way of building models
which are quantum group invariant uses the TL algebra. It is a algebra
generated by the Hamiltonian density }$U_k${\sf \ , }$k=1,2,...,N-1${\sf \
subject to the following constraints } 
\begin{equation}
\begin{array}{lll}
U_k^2=(Q+Q^{-1})U_k & , & U_kU_{k\pm 1}U_k=U_k \\ 
&  &  \\ 
U_kU_j=U_jU_k &  & |k-j|>1.
\end{array}
\label{int1}
\end{equation}
{\sf with }$Q\in C${\sf \ a given number. This algebra appears in a large
class of quantum lattice models and leads, at level of free and ground state
energies, to some equivalence among the models \cite{B}.}

{\sf Taking into account usual boundary conditions, the TL Hamiltonians take
the form } 
\begin{equation}
{\cal H}=\sum_{k=1}^{N-1}U_{kk+1}+{\bf bt}  \label{pbah}
\end{equation}
{\sf where }$U_{kk+1}\equiv U_k${\sf \ operates in a direct product of
complex spaces at positions }$k${\sf \ and }$k+1${\sf . In general, they are
not invariant with respect to quantum groups since the boundary terms, }$%
{\bf bt}${\sf , break translational invariance, reflecting the
non-cocommutativity of the co-product. Indeed, we know from \cite{Alcaraz,
PS, RM} that very special boundary terms must be considered when we seek
these quantum group invariant spin chains. In particular, one possibility is
to consider free boundary conditions, i.e., }${\bf bt}=0${\sf \ . For The
XXZ-Hamiltonian with free boundary conditions one has to apply the Bethe
ansatz \cite{Bethe} techniques introduced first by Alcaraz et al \cite
{Alcaraz} and after by Sklyanin \cite{Sklyanin} using Cherednik's reflection
matrices \cite{Cherednik, MN}. By this method the XXZ-Heisenberg model \cite
{Destri}, the }$spl_q(2,1)${\sf \ invariant supersymmetric t-J model \cite
{Angela, Ruiz} , the }$U_q[sl(n)]${\sf \ invariant generalization of the
XXZ-chain \cite{Vega} and the }$SU_q(n|m)${\sf \ spin chains \cite{RM, Yue}
have been solved for free boundary conditions.}

{\sf Recently, by means of a generalized algebraic nested Bethe ansatz,
Karowski and Zapletal \cite{Karowski} presented a class of quantum group
invariant }$n${\sf -state vertex models with periodic boundary conditions.
Also an extension of this method to the case of graded vertex models was
analyzed in \cite{Angela2}, where a }$spl_q(2|1)${\sf \ invariant susy }$t$%
{\sf -}$J${\sf \ model with closed boundary conditions was presented.}

{\sf In fact, these kind of models were first discussed by Martin \cite
{Martin} from the representations of the Hecke algebra. In this case, the
boundary term is a non-local operator, }${\bf bt}={\cal U}_0${\sf \ (see
Section }$4${\sf ). The study of closed quantum group invariant closed spin
chains in the framework of the coordinate Bethe ansatz was presented by
Grosse et al \cite{Pallua} for the fundamental representation of }$SU_q(2)$%
{\sf \ and generalized by \cite{LG}. In this context it would be interesting
to discuss other quantum group invariant closed spin chains.}

{\sf In the present paper we find the spectrum of Hamiltonians based on
representations of the TL algebra associated with quantum groups. The Bethe
ansatz equations for different types of boundary conditions (periodic,
closed and free) are obtained through a modified version of the coordinate
Bethe ansatz. Therefore we generalize the results of ref.\cite{LG}.Our
method was recently applied to solve graded T-L Hamiltonian \cite{L} and
anisotropic correlated electron model associated with this algebra \cite{LRA}%
.}

{\sf The paper is organized as follows. In Section }$2${\sf , we describe
the representations of the TL algebra, constructed as projectors on total
spin zero of two neighboring spins. In Section }$3${\sf , we introduce the
modified coordinate Bethe Ansatz for the TL Hamiltonians with periodic
boundary conditions. In Section }$4${\sf \ their Bethe Ansatz solution is
presented with non-local boundary conditions. In section }$5${\sf , contains
the solution for free boundary conditions. Finally the conclusions are
reserved for section }$6${\sf .}

\section{\sf Representations of the TL algebra}

{\sf Representations of the TL algebra, commuting with quantum groups, can
be constructed in the following way \cite{R}. Suppose }${\cal U}_q(X_n)${\sf %
\ is the universal enveloping algebra of a finite dimensional Lie algebra }$%
X_n${\sf , equipped with the coproduct }$\Delta :{\cal U}_q\rightarrow {\cal %
U}_q\otimes {\cal U}_q${\sf \ \cite{D}. If now }$\pi :{\cal U}_q\rightarrow 
{\rm En{\sf d}\ }V_\Lambda ${\sf \ is a finite dimensional irreducible
representation with highest weight }${\Lambda }${\sf \ and we assume that
the decomposition }$V_\Lambda \otimes V_\Lambda ${\sf \ is multiplicity free
and includes one trivial representation on }$V_0${\sf , then the projector }$%
{\cal P}_0${\sf \ from }$V_\Lambda \otimes V_\Lambda ${\sf \ onto }$V_0${\sf %
\ is a representation of the TL algebra. The deformation parameter }$q${\sf %
, which plays the role of a coupling constant in the Hamiltonian, is related
to }$Q${\sf \ as: } 
\begin{equation}
Q+Q^{-1}={\rm Tr}_{V_\Lambda }(q^{-2\rho }),  \label{int3}
\end{equation}
{\sf where }$\rho ${\sf \ is half the sum of the positive roots.}

{\sf We will consider the following specific cases, }$(V_\Lambda ,{\cal U}%
_q(X_n))=(V_{2s\Lambda _1},{\cal U}_q(A_1))${\sf \ for spin }$s${\sf , }$%
(V_{\Lambda _1},{\cal U}_q(B_n)${\sf ,}$(V_{\Lambda _1},{\cal U}_q(C_n)${\sf %
\ and }$(V_{\Lambda _1},{\cal U}_q(D_n)${\sf . Namely we treat the }$q${\sf %
-deformations of the spin-}$s${\sf \ representation of }$sl(2)${\sf \ and
the vector representations of }$so(2n+1),sp(2n)${\sf \ and }$so(2n)${\sf . }$%
V_\Lambda ${\sf \ denotes the }${\cal U}_q(X_n)${\sf \ module with highest
weight }$\Lambda ${\sf \ and }$\Lambda _1${\sf \ is a highest weight of }$%
X_n ${\sf .}

{\sf Since we are not going to use any group-theoretical machinery, we will
just lift the relevant formulas off Batchelor and Kuniba \cite{BK} in order
to display explicitly the Hamiltonians to be diagonalized.}

{\sf We introduce orthonormal vectors }$e_\mu ,${\sf \ }$<e_\mu ,e_\nu
>=\delta _{\mu \nu }${\sf , to express }$\rho ${\sf \ and the set }${\cal A}$%
{\sf \ of weights appearing in the representation }$\pi ${\sf \ of }${\cal U}%
_q(X_n)${\sf \ as follows:}

\begin{itemize}
\item[$A_{1}$]  {\sf : Spin-}$s${\sf \ representation of the }$sl(2)${\sf \
algebra } 
\begin{eqnarray}
\ \left. {\cal A}=\{se_{12},(s-1)e_{12},\ldots ,-se_{12}\},\quad \rho =\frac
12e_{12},\quad (e_{12}=e_1-e_2),\right.  \nonumber \\
\ \left. J=\{s,s-1,\ldots ,-s\},\quad \epsilon (\mu )=(-1)^{\tilde \mu
},\quad Q+Q^{-1}=[2s+1].\right.  \label{int4}
\end{eqnarray}

\item[$B_{n}$]  {\sf : Vector representation of the }$so(2n+1)${\sf \
algebra }$(n\geq 2)$%
\begin{eqnarray}
\ \left. {\cal A}=\{0,\pm e_1,\ldots ,\pm e_n\},\ \rho =\sum_{\alpha
=1}^n(n-(\alpha -\frac 12))e_\alpha ,\ J=\{0,\pm 1,\ldots ,\pm n\},\right. 
\nonumber \\
\ \left. \epsilon (\mu )=(-1)^{\tilde \mu },\quad Q+Q^{-1}=\frac{%
[2n-1][n+1/2]}{[n-1/2]}\ .\right.  \label{int5}
\end{eqnarray}

\item[$C_{n}$]  {\sf : Vector representation of the }$sp(2n)${\sf \ algebra }%
$(n\geq 1)$%
\begin{eqnarray}
\ \left. {\cal A}=\{\pm e_1,\ldots ,\pm e_n\},\quad \rho =\sum_{\alpha
=1}^n(n-\alpha +1)e_\alpha ,\quad J=\{\pm 1,\ldots ,\pm n\},\right. 
\nonumber \\
\ \qquad \left. \epsilon (\mu )={\rm sign}(\mu ),\quad Q+Q^{-1}=\frac{%
[n][2n+2]}{[n+1]}\ .\right.  \label{int6}
\end{eqnarray}

\item[$D_{n}$]  {\sf : Vector representations of the }$so(2n)${\sf \ algebra 
}$(n\geq 3)$%
\begin{eqnarray}
\ \left. {\cal A}=\{\pm e_1,\ldots ,\pm e_n\},\quad \rho =\sum_{\alpha
=1}^{n-1}(n-\alpha )e_\alpha ,\quad J=\{\pm 1,\ldots ,\pm n\},\right. \qquad
\nonumber \\
\ \left. \epsilon (\mu )=1,\quad Q+Q^{-1}=\frac{[2n-2][n]}{[n-1]}\ .\right.
\label{int7}
\end{eqnarray}
\end{itemize}

{\sf For }$\mu \in J${\sf \ the symbol }$\tilde \mu ${\sf \ is defined as }$%
\tilde \mu =\mu +1/2${\sf \ for }$A_1${\sf \ with }$s${\sf \ semi-integer
and }$\tilde \mu =\mu ${\sf \ for }$A_1${\sf \ with }$s${\sf \ integer. For }%
$B_n${\sf , }$C_n${\sf \ and }$D_n${\sf , }$\tilde \mu =0${\sf \ with the
exception of }$\tilde 0=1${\sf \ for }$B_n${\sf . The }$q${\sf -number
notation is }$[x]=(q^x-q^{-x})/(q-q^{-1})${\sf .}

{\sf For }$X_n=B_n,C_n${\sf \ and }$D_n${\sf , we extend the suffix of }$%
e_\mu ${\sf \ to }$-n\leq \mu \leq n${\sf \ by setting }$e_{-\mu }=-e_\mu $%
{\sf \ (hence }$e_0=0${\sf ). Using the index set }$J${\sf \ above, we can
write }${\cal A}=\{\mu (e_1-e_2)\}${\sf \ for }$A_1${\sf \ and }${\cal A}%
=\{e_\mu |\mu \in J)\}${\sf \ for }$B_n,C_n${\sf \ and }$D_n${\sf .}

{\sf Denoting by }$E_{\mu \nu }\in {\rm End}\,V_\Lambda ${\sf \ the matrix
unit, having all elements zero, except at row }$\mu ${\sf \ and column }$\nu 
${\sf , the projector can be written as } 
\begin{equation}
{\cal P}_0=\frac 1{Q+Q^{-1}}\sum_{\mu ,\nu \in J}\epsilon (\mu )\epsilon
(\nu )q^{-<e_\mu +e_\nu ,\rho >}E_{\mu \nu }\otimes E_{-\mu -\nu }.
\label{int8}
\end{equation}

{\sf In the following we will refer to all models generically as higher spin
models for simplicity, even when not talking about }$A_1${\sf . The
dimensions of }$B_n${\sf \ are related with the dimensions of }$A_1${\sf \
for }$s${\sf \ integer and the dimensions of }$C_n${\sf \ and }$D_n${\sf \
are related with the dimensions of }$A_1${\sf \ for }$s${\sf \ semi-integer.}

{\sf If we consider then a one-dimensional chain of length }$N${\sf \ with a
''spin'' at each site, the ''spin variables '' range over the set of weight
vectors }${\cal A}=\left\{ v_\mu |\mu \in J\right\} ${\sf \ and our Hilbert
space is an }$N${\sf -fold tensor product }$V_\Lambda \otimes \ldots \otimes
V_\Lambda ${\sf . For }$A_1${\sf , these are the }$q${\sf -analogs of the
usual spin states. The periodic Hamiltonians associated with the TL
representations are given by the following sum over }$N${\sf \ sites } 
\begin{equation}
{\cal H}=\sum_{k=1}^NU_k.  \label{int2}
\end{equation}
{\sf Here }${\bf bt}=U_{N,N+1}=U_{N1}${\sf \ and the Hamiltonian densities
acting on two neighboring sites are then given by: } 
\begin{eqnarray}
\langle \nu ,\lambda |U|\mu ,\kappa \rangle &=&\epsilon (\mu )\epsilon (\nu
)q^{-<e_\mu +e_\nu ,\rho >}\delta _{\mu +\kappa ,0}\,\delta _{\nu +\lambda
,0}  \nonumber \\
\mu ,\nu ,\kappa ,\lambda &\in &J  \label{int9}
\end{eqnarray}

{\sf Using representations of the TL algebra, one can also build solvable
vertex models whose Hamiltonian limit leads to the previously mentioned
quantum spin chains. To do so, we introduce an operator }$R(u)\in {\rm End\ }%
(V_\Lambda \otimes V_\Lambda )${\sf \ by } 
\begin{eqnarray*}
R_k(u) &=&I\otimes \cdots \otimes I\otimes {%
\raisebox{-1.3em}
{$\stackrel{\textstyle \underbrace{R(u)}} {\scriptstyle k,k+1}$}}\otimes
I\otimes \cdots \otimes I \\
R(u) &=&\sum_{\mu ,\nu ,\kappa ,\lambda \in J}R_{\mu \kappa }^{\nu \lambda
}(u,\eta )E_{\mu \nu }^{(k)}\otimes E_{\kappa \lambda }^{(k+1)}
\end{eqnarray*}
\begin{equation}
\left. R_{\mu \kappa }^{\nu \lambda }(u,\eta )=\frac{\sinh (\eta -u)}{\sinh
\eta }\delta _{\mu \nu }\delta _{\kappa \lambda }+\frac{\sinh u}{\sinh \eta }%
\epsilon (\mu )\epsilon (\nu )q^{-<e_\mu +e_\nu ,\rho >}\delta _{\mu +\kappa
,0}\,\delta _{\nu +\lambda ,0}\right.  \label{int10}
\end{equation}
{\sf where }$u${\sf \ is the spectral parameter and the anisotropy parameter 
}$\eta ${\sf \ is chosen so that } 
\begin{equation}
2\cosh \eta =Q+Q^{-1}  \label{int11}
\end{equation}
{\sf The }$R${\sf \ matrix commutes with the quantum group action and the
Yang-Baxter equation } 
\begin{equation}
R_k(u)R_{k+1}(u+v)R_k(v)=R_{k+1}(v)R_k(u+v)R_{k+1}(u)  \label{int12}
\end{equation}
{\sf is valid owing to the TL relations (\ref{int1}).}

{\sf Due to the TL algebra these models are equivalent to the }$6${\sf %
-vertex and the '' }$p${\sf -state '' self-dual Potts models ( }$\sqrt{p}%
=\lim_{q\rightarrow 1}(Q+Q^{-1})${\sf ) through the argument in \cite{B}. In
fact, the cases }${\cal U}_q(A_1)${\sf \ with }$s=1/2${\sf \ and }${\cal U}%
_q(C_1)${\sf \ yield the }$6${\sf -vertex model itself. When }$q=1${\sf ,
the vertex models here reduce to those discussed in \cite{PKS, TT, PW},
where the number of states is equal to }$\sqrt{p}${\sf .}

{\sf The case }$A_1${\sf \ have been studied by several authors. When }$q=1$%
{\sf \ and }$s=1/2${\sf \ the operator }$U_k${\sf \ is essentially the
Heisenberg interaction term }$\sigma _k^x\sigma _{k+1}^x+\sigma _k^y\sigma
_{k+1}^y+\sigma _k^z\sigma _{k+1}^z${\sf . Besides the explicit matrix
elements , available from (\ref{int9}), the local Hamiltonian }$U_k${\sf \
is in principle also expressible in terms of the usual representation
matrices for }$SU(2)${\sf \ generators. The resulting }$q${\sf -deformed
Hamiltonian has been written down for }$s=1${\sf \ in \cite{BNMR}. However
for }$s>1${\sf , writing the Hamiltonian in terms of the usual }$SU(2)${\sf %
\ operators becomes very cumbersome. These Hamiltonians have alternative
expressions in terms of Casimir operators \cite{BK}.}

{\sf The limit }$q\rightarrow 1${\sf \ has discussed for general }$s${\sf \ 
\cite{KP, BB}. As for related spin-}$1${\sf \ biquadratic model \cite{KP,
BB1}, they are massive for }$s\geq 1${\sf \ and of relevance to the
dimerization transition on }$SU(n)${\sf \ antiferromagnetic chains \cite{A}.
The case }$s=1/2${\sf \ has also been investigated in some detail in \cite
{PS}.}

{\sf Having now built common ground for all models, whose salient feature is
they being spin zero projectors, we may now follow the steps of reference 
\cite{KL} to find their spectra.}

\section{\sf Bethe Ansatz: Periodic boundary conditions}

{\sf All the above Hamiltonians are }$U(1)${\sf \ invariant and we can
classify their spectra according to sectors. For }$A_1${\sf \ the commuting
operator is the total spin }$S^z=\sum_{k=1}^NS_k^z${\sf \ where }$S_k^z=$%
{\sf diag}$(s,s-1,...,-s+1,-s)_k${\sf \ and we set the conserved quantum
number }$r=sN-S^z${\sf . We extend these quantum numbers for the other
algebras as }$r=N\omega -{\bf S}^z${\sf \ where }${\bf S}^z=\sum_k{\bf S}%
_k^z ${\sf \ with }${\bf S}_k^z=${\sf diag}$(\omega ,\omega -1,...,-\omega
+1,-\omega )_k${\sf \ and }$\omega ={\rm max}\{J\}${\sf . Eigenvalues of the
operator }$r${\sf \ can be used to collect the eigenstates of }${\cal H}$%
{\sf \ in sectors, }$\Psi _r${\sf . Therefore, there exists a reference
state }$\Psi _0${\sf , satisfying }${\cal H}\Psi _0=E_0\Psi _0${\sf , with }$%
E_0=0${\sf . We take }$\Psi _0${\sf \ to be }$\Psi _0=|\omega \ \omega \
\omega \cdots \omega \rangle ${\sf . It is the only eigenstate in the sector 
}$r=0${\sf \ and all other energies will be measured relative to this state.}

{\sf In every sector }$r${\sf \ there are eigenstates degenerate with }$\Psi
_0${\sf . They contain a set of impurities. We call impurity any state
obtained by lowering some of the }$\left| \omega ,k\right\rangle ${\sf 's,
such that the sum of any two neighboring spins is non-zero. Since }${\cal H}$%
{\sf \ is a projector on spin zero, all these states are annihilated by }$%
{\cal H}${\sf . In particular, they do not move under the action of }${\cal H%
}${\sf , which is the reason for their name.}

{\sf Nothing interesting happens in sectors with }$r<2\omega ${\sf . Sector }%
$r=2\omega ${\sf , we encounter the situation where the states }$\left|
\alpha ,k\right\rangle ${\sf \ and }$\left| -\alpha ,k\pm 1\right\rangle $%
{\sf , }$\alpha \in J${\sf , occur in neighboring pairs. They move under the
action of }${\cal H}${\sf , i.e., the sector }$r=2\omega ${\sf \ contains
one free pseudoparticle. In general, for a sector }$r${\sf \ we may have }$p$%
{\sf \ pseudoparticles and }$N_{\omega -1},N_{\omega -2},...,N_{-\omega +1}$%
{\sf \ impurities of the type }$\omega -1,\omega -2,...,-\omega +1,${\sf \
respectively, such that } 
\begin{equation}
r=2\omega p+\sum_{\alpha =1}^{2\omega -1}\alpha N_{\omega -\alpha }.
\label{pba1}
\end{equation}

{\sf The main result of this section is to show that }${\cal H}${\sf \ can
be diagonalized in a convenient basis, constructed from products of single
pseudoparticle wavefunctions. The energy eigenvalues will be parametrized as
a sum of single pseudoparticle contributions.}

{\sf The first nontrivial sector is }$r=2\omega ${\sf \ and the
correspondent eigenspace is spanned by the states }$|k(-\alpha ,\alpha
)>=|\omega \ \omega \cdots \omega \ {%
\raisebox{-0.65em}{$\stackrel{\textstyle {\scriptstyle{-}}\alpha}
{\scriptstyle k}$}\ }\alpha \ \omega \cdots \omega >${\sf \ , where }$%
k=1,2,...,N-1\ ${\sf and\ }$\ \alpha \in J.${\sf \ We seek eigenstates of }$%
{\cal H}${\sf \ which are linear combinations of these vectors. It is very
convenient to consider the linear combination } 
\begin{equation}
\left| \Omega (k)\right\rangle =\sum_{\alpha =-\omega }^\omega \epsilon
(\omega )\epsilon (\alpha )\ q^{-<e_\omega +e_\alpha ,\rho >}\left|
k(-\alpha ,\alpha )\right\rangle ,  \label{pba2}
\end{equation}
{\sf which is an eigenstate of }$U_k${\sf : } 
\begin{equation}
U_k\left| \Omega (k)\right\rangle =(Q+Q^{-1})\left| \Omega (k)\right\rangle .
\label{pba3}
\end{equation}
{\sf Moreover, the action of }$U_{k\pm 1}${\sf \ on }$\left| \Omega
(k)\right\rangle ${\sf \ is very simple } 
\begin{equation}
\begin{array}{lll}
U_{k-1}\left| \Omega (k)\right\rangle =\epsilon \left| \Omega
(k-1)\right\rangle &  & U_{k+1}\left| \Omega (k)\right\rangle =\epsilon
\left| \Omega (k+1)\right\rangle \\ 
&  &  \\ 
U_k\left| \Omega (m)\right\rangle =0 &  & k\neq \{m\pm 1,m\}
\end{array}
\label{pba4}
\end{equation}
{\sf where }$\epsilon =1${\sf \ for }$B_n${\sf , }$D_n${\sf \ and }$A_1${\sf %
\ ( }$s${\sf \ integer) and }$\epsilon =-1${\sf \ for }$C_n${\sf \ and }$A_1$%
{\sf \ ( }$s${\sf \ semi-integer).}

{\sf It should be emphasized that although the linear combination (\ref{pba2}%
) is different for each model, the action of }$U_k${\sf \ is always given by
(\ref{pba3}) and (\ref{pba4}). Therefore, all Hamiltonians can be treated in
a similar way and it affords a considerable simplification in their
diagonalizations when we compare with the calculus used in the usual spin
basis \cite{KL}.}

\subsection{\sf One-pseudoparticle}

{\sf We will now start to diagonalize }${\cal H}${\sf \ in every sector. Let
us consider one free pseudoparticle as a highest weight state which lies in
the sector }$r=2\omega $

\begin{equation}
\Psi _{2\omega }=\sum_kA(k)\left| \Omega (k)\right\rangle .  \label{pba5}
\end{equation}
{\sf Using the eigenvalue equation }${\cal H}${\sf \ }$\Psi _{2\omega
}=E_{2\omega }\Psi _{2\omega }${\sf , one can derive a complete set of
equations for the wavefunctions }$A(k)${\sf .}

{\sf When the bulk of }${\cal H}${\sf \ acts on }$\left| \Omega
(k)\right\rangle ${\sf \ it sees the reference configuration, except in the
vicinity of }$k${\sf \ where we use (\ref{pba3}) and (\ref{pba4}) to get } 
\begin{eqnarray}
{\cal H}\left| \Omega (k)\right\rangle =(Q+Q^{-1})\left| \Omega
(k)\right\rangle +\epsilon \left| \Omega (k-1)\right\rangle +\epsilon \left|
\Omega (k+1)\right\rangle  \nonumber \\
2\leq k\leq N-2  \label{pba6}
\end{eqnarray}
{\sf Substituting (\ref{pba6}) in the eigenvalue equation, we get } 
\begin{eqnarray}
(E_{2\omega }-Q-Q^{-1})A(k)=\epsilon A(k-1)+\epsilon A(k+1)  \nonumber \\
2\leq k\leq N-2  \label{pba7}
\end{eqnarray}
{\sf Here we will treat periodic boundary conditions . They demand }$%
U_{N,N+1}=U_{N,1}${\sf , implying }$A(k+N)=A(k)${\sf . This permits us to
complete the set of equations (\ref{pba7}) for }$A(k)${\sf \ by including
the equations for }$k=1${\sf \ and }$k=N-1${\sf . Now we parametrize }$A(k)$%
{\sf \ by plane wave }$A(k)=A\xi ^k${\sf \ to get the energy of one free
pseudoparticle as: } 
\begin{eqnarray}
E_{2\omega }=Q+Q^{-1}+\epsilon \left( \xi +\xi ^{-1}\right)  \nonumber \\
\xi ^N=1  \label{pba8}
\end{eqnarray}
{\sf Here }$\xi ={\rm e}^{i\theta }${\sf , }$\theta ${\sf \ being the
momenta determined from the periodic boundary to be }$\theta =2\pi l/N${\sf %
, with }$l${\sf \ an integer.}

\subsection{\sf One-pseudoparticle and impurities}

{\sf Let us consider the state with one pseudoparticle and one impurity of
type }$(\omega -1)${\sf , which lies in the sector }$r=2\omega +1${\sf . We
seek eigenstates in the form}

\begin{equation}
\Psi _{2\omega +1}(\xi _1,\xi _2)=\sum_{k_1<k_2}\left\{ A_1(k_1,k_2)\left|
\Omega _1(k_1,k_2)\right\rangle +A_2(k_1,k_2)\left| \Omega
_2(k_1,k_2)\right\rangle \right\}  \label{pba9}
\end{equation}
{\sf We try to build these eigenstates out of translational invariant
products of one pseudoparticle excitation with parameter }$\xi _2${\sf \ and
one impurity with parameter }$\xi _1${\sf : } 
\begin{equation}
\Psi _{2\omega +1}(\xi _1,\xi _2)=\left| (\omega -1)(\xi _1)\right\rangle
\times \Psi _{2\omega }(\xi _2)+\Psi _{2\omega }(\xi _2)\times \left|
(\omega -1)(\xi _1)\right\rangle  \label{pba9a}
\end{equation}
{\sf Using one-pseudopaticle eigenstate solution (\ref{pba5}) and comparing
this with (\ref{pba9}) we get } 
\begin{eqnarray}
\left| \Omega _1(k_1,k_2)\right\rangle &=&\sum_{\alpha =-\omega }^\omega
\epsilon (\omega )\epsilon (\alpha )\ q^{-<e_\omega +e_\alpha ,\rho >}\left|
k_1(\omega -1),k_2(-\alpha ,\alpha )\right\rangle  \nonumber \\
&&  \nonumber \\
\left| \Omega _2(k_1,k_2)\right\rangle &=&\sum_{\alpha =-\omega }^\omega
\epsilon (\omega )\epsilon (\alpha )\ q^{-<e_\omega +e_\alpha ,\rho >}\left|
k_1(-\alpha ,\alpha ),k_2(\omega -1)\right\rangle  \nonumber \\
&&  \label{pba10}
\end{eqnarray}
{\sf and } 
\begin{equation}
A_1(k_1,k_2)=A_1\xi _1^{k_1}\xi _2^{k_2}\qquad ,\qquad A_2(k_1,k_2)=A_2\xi
_2^{k_1}\xi _1^{k_2}.  \label{pba12}
\end{equation}
{\sf Periodic boundary conditions }$A_1(k_2,N+k_1)=A_2(k_1,k_2)${\sf \ and }$%
A_i(N+k_1,N+k_2)=A_i(k_1,k_2)${\sf ,\quad }$i=1,2${\sf \ imply that } 
\begin{equation}
A_1\xi _2^N=A_2\quad ,\quad \xi ^N=(\xi _1\xi _2)^N=1\   \label{pba13}
\end{equation}

{\sf When }${\cal H}${\sf \ now acts on }$\Psi _{2\omega +1}${\sf , we will
get a set of coupled equations for }$A_i(k_1,k_2),${\sf \ }$i=1,2${\sf . We
split the equations into far equations, when the pseudoparticle do not meet
the impurity and near equations, containing terms when they are neighbors.}

{\sf Since the impurity is annihilated by }${\cal H}${\sf , the action of }$%
{\cal H}${\sf \ on (\ref{pba9}) in the case far (i.e., }$(k_2-k_1)\geq 3$%
{\sf ), can be write down directly from (\ref{pba7}) :}

\begin{eqnarray}
\left( E_{2\omega +1}-Q-Q^{-1}\right) A_1(k_1,k_2)=\epsilon
A_1(k_1,k_2-1)+\epsilon A_1(k_1,k_2+1)  \label{pba14} \\
\left( E_{2\omega +1}-Q-Q^{-1}\right) A_2(k_1,k_2)=\epsilon
A_2(k_1-1,k_2)+\epsilon A_2(k_1+1,k_2)  \label{pba15}
\end{eqnarray}
{\sf Using the parametrization (\ref{pba12}), these equations will give us
the energy eigenvalues } 
\begin{equation}
E_{2\omega +1}=Q+Q^{-1}+\epsilon (\xi _2+\xi _2^{-1})  \label{pba16}
\end{equation}
{\sf To find }$\xi _2${\sf \ we must consider the\ near equations. First, we
compute the action of }${\cal H}${\sf \ on the coupled near states }$\left|
\Omega _1(k,k+1)\right\rangle ${\sf \ and }$\left| \Omega
_2(k,k+2)\right\rangle ${\sf :}

\begin{eqnarray}
{\cal H}\left| \Omega _1(k,k+1)\right\rangle =(Q+Q^{-1})\left| \Omega
_1(k,k+1)\right\rangle +\epsilon \left| \Omega _1(k,k+2)\right\rangle
+\epsilon \left| \Omega _2(k,k+2)\right\rangle  \nonumber \\
&&  \label{pba17} \\
{\cal H}\left| \Omega _2(k,k+2)\right\rangle =(Q+Q^{-1})\left| \Omega
_2(k,k+2)\right\rangle +\epsilon \left| \Omega _2(k-1,k+2)\right\rangle
+\epsilon \left| \Omega _1(k,k+1)\right\rangle  \nonumber \\
&&  \label{pba18}
\end{eqnarray}
{\sf The last terms in these equations tell us that a pseudoparticle can
propagate past the isolated impurity, but in so doing causes a shift in its
position by two lattice site. Substituting (\ref{pba17}) and (\ref{pba18})
into the eigenvalue equation, we get}

\begin{eqnarray}
\left( E_{2\omega +1}-Q-Q^{-1}\right) A_1(k,k+1)=\epsilon
A_1(k,k+2)+\epsilon A_2(k,k+2)  \label{pba19} \\
\left( E_{2\omega +1}-Q-Q^{-1}\right) A_2(k,k+2)=\epsilon
A_2(k-1,k+2)+\epsilon A_1(k,k+1)  \label{pba20}
\end{eqnarray}
{\sf These equations, which are not automatically satisfied by the ansatz (%
\ref{pba12}), are equivalent to the conditions } 
\begin{equation}
A_1(k,k)\equiv A_2(k,k+2)\qquad ,\qquad A_1(k,k+1)\equiv A_2(k+1,k+2).
\label{pba21}
\end{equation}
{\sf obtained by subtracting Eq. (\ref{pba19}) from Eq.(\ref{pba14}) for }$%
k_1=k${\sf \ }$,${\sf \ }$k_2=k+1${\sf \ and by subtracting Eq. (\ref{pba20}%
) from Eq.(\ref{pba15}) for }$k_1=k${\sf \ }$,${\sf \ }$k_2=k+2${\sf ,
respectively. The conditions (\ref{pba21}) require a modification of the
amplitude relation (\ref{pba13}): } 
\begin{equation}
\frac{A_2}{A_1}=\xi _1^{-2}=\xi _2^N\Rightarrow \xi _2^N\xi _1^2=1\qquad {%
{\rm or}\qquad }\xi _2^{N-2}\xi ^2=1  \label{pba22}
\end{equation}
{\sf Putting }$\xi _i={\rm e}^{i\theta _i}${\sf , }$i=1,2${\sf , it means }$%
\cos (N-2)\theta _2=\cos 2\theta _1${\sf . Hence}

\begin{equation}
\theta _2=\frac{2\pi m\pm 4\pi l/N}{N-2},\qquad l\ {\rm and\ }m\ {\rm %
integers}{\rm .}  \label{pba23}
\end{equation}
{\sf In other words, }$\Psi _{2\omega +1}(\xi _1,\xi _2)${\sf \ are
eigenstates of }${\cal H}${\sf \ with energy eigenvalues given by }$%
E_{2\omega +1}=Q+Q^{-1}+2\epsilon \cos \theta _2${\sf . \ Note that when }$%
Nm\pm 2l${\sf \ is a multiple of }$(N-2)${\sf \ we get states which are
degenerate with the one-pseudoparticle states }$\Psi _{2\omega }${\sf ,
which lie in the sector }$r=2\omega ${\sf .}

{\sf In the sectors }$r=2\omega +l${\sf \ we also will find states, which
consist of one pseudoparticle with parameter }$\xi _{l+1}${\sf \ interacting
with }$l${\sf \ impurities, distributing according to (\ref{pba1}), with
parameters }$\xi _i,i=1,2...,l${\sf .}

{\sf The energy of these states is parametrized as in (\ref{pba16}) and }$%
\xi _{l+1}${\sf \ satisfies the condition (\ref{pba22}) with }$\xi =\xi
_1\cdots \xi _l\ \xi _{l+1}${\sf . It involves only }$\xi _{l+1}${\sf \ and }%
$\xi _{{\rm imp}}=\xi _1\ \xi _2\cdots \xi _l${\sf , being therefore highly
degenerate, i.e. } 
\begin{equation}
\xi _{l+1}^N\xi _1^2\ \xi _2^2\cdots \xi _l^2=1  \label{pba24}
\end{equation}
{\sf This is to be expected due to the irrelevance of the relative
distances, up to jumps of two positions via exchange with a pseudoparticle.
Moreover, these results do not depend on impurity type.}

\subsection{\sf Two-pseudoparticles}

{\sf The sector }$r=4\omega ${\sf \ contains, in addition to the cases
discussed above, states which consist of two interacting pseudoparticles. We
seek eigenstates in the form } 
\begin{equation}
\Psi _{4\omega }(\xi _1,\xi _2)=\sum_{k_1+1<k_2}A(k_1,k_2)\left| \Omega
(k_1,k_2)\right\rangle  \label{pba25}
\end{equation}
{\sf Again, we try to build two-pseudoparticle eigenstates out of
translational invariant products of one-pseudoparticle excitations at }$k_1$%
{\sf \ and }$k_2${\sf \ (}$k_2${\sf \ }$\geq k_1${\sf \ }$+2${\sf ) : } 
\begin{equation}
\Psi _{4\omega }(\xi _1,\xi _2)=\Psi _{2\omega }(\xi _1)\times \Psi
_{2\omega }(\xi _2)+\Psi _{2\omega }(\xi _2)\times \Psi _{2\omega }(\xi _1)
\label{pba26}
\end{equation}
{\sf Using again (\ref{pba5}) and comparing (\ref{pba26}) with (\ref{pba25})
we get } 
\begin{equation}
\left| \Omega (k_1,k_2)\right\rangle =\sum_{\alpha ,\ \beta =-\omega
}^\omega \epsilon (\alpha )\epsilon (\beta )\ q^{-<2e_\omega +e_\alpha
+e_\beta ,\rho >}\left| k_1(-\alpha ,\alpha ),k_2(-\beta ,\beta
)\right\rangle  \label{pba27}
\end{equation}
{\sf for }$k_2\geq k_1+3${\sf \ and } 
\begin{equation}
A(k_1,k_2)=A_{12}\xi _1^{k_1}\xi _2^{k_2}+A_{21}\xi _2^{k_1}\xi _1^{k_2},
\label{pba28}
\end{equation}
{\sf for }$k_2${\sf \ }$\geq k_1${\sf \ }$+2.$

{\sf Periodic boundary conditions }$A(k_2,N+k_1)=A(k_1,k_2)${\sf \ and }$%
A(N+k_1,N+k_2)=A(k_1,k_2)${\sf \ imply } 
\begin{equation}
A_{12}\xi _2^N=A_{21}\qquad {\rm and}\qquad \xi ^N=1  \label{pba29}
\end{equation}
{\sf where }$\xi =\xi _1\xi _2${\sf \ (}$\xi _i={\rm e}^{i\theta _i},\ i=1,2$%
{\sf ) and the total momentum is }$\theta _1+\theta _2=2\pi l/N${\sf , with }%
$l${\sf \ integer.}

{\sf Applying }${\cal H}${\sf \ to the state of (\ref{pba25}), we obtain a
set of equations for the wavefunctions }$A(k_1,k_2)${\sf . When the two
pseudoparticles are separated, (}$k_2\geq k_1+3${\sf ) these are the
following far equations: } 
\begin{equation}
\begin{array}{lll}
\left( E_{4\omega }-2Q-2Q^{-1}\right) A(k_1,k_2) & = & \epsilon
A(k_1-1,k_2)+\epsilon A(k_1+1,k_2) \\ 
&  &  \\ 
& + & \epsilon A(k_1,k_2-1)+\epsilon A(k_1,k_2+1)
\end{array}
\label{pba30}
\end{equation}
{\sf We already know them to be satisfied, if we parametrize }$A(k_1,k_2)$%
{\sf \ by plane waves (\ref{pba28}). The corresponding energy eigenvalue is }
\begin{equation}
E_{4\omega }=2Q+2Q^{-1}+\epsilon \left( \xi _1+\xi _1^{-1}+\xi _2+\xi
_2^{-1}\right)  \label{pba31}
\end{equation}

{\sf The real problem arises of course, when pseudoparticles are neighbors,
so that they interact and we have no guarantee that the total energy is sum
of single pseudoparticle energies.}

{\sf Acting of }${\cal H}${\sf \ on the state (\ref{pba27}) gives the
following set of equations for the near states } 
\begin{equation}
\begin{array}{lll}
{\cal H}\left| \Omega (k,k+2)\right\rangle & = & 2\left( Q+Q^{-1}\right)
\left| \Omega (k,k+2)\right\rangle +\epsilon \left| \Omega
(k-1,k+2)\right\rangle \\ 
&  &  \\ 
& + & \epsilon \left| \Omega (k,k+3)\right\rangle +U_{k+1}\left| \Omega
(k,k+2)\right\rangle
\end{array}
\label{pba32}
\end{equation}

{\sf Before we substitute this result into the eigenvalue equation, we
observe that some new states are appearing. In order to incorporate these
new states in the eigenvalue problem, we define } 
\begin{equation}
U_{k+1}\left| \Omega (k,k+2)\right\rangle \equiv \epsilon \left| \Omega
(k,k+1)\right\rangle +\epsilon \left| \Omega (k+1,k+2)\right\rangle
\label{pba33}
\end{equation}
{\sf Here we underline that we are using the same notation for these new
states. Applying }${\cal H}${\sf \ to them we obtain } 
\begin{equation}
\begin{array}{lll}
{\cal H}\left| \Omega (k,k+1)\right\rangle & = & \left( Q+Q^{-1}\right)
\left| \Omega (k,k+1)\right\rangle +\epsilon \left| \Omega
(k-1,k+1)\right\rangle \\ 
&  &  \\ 
& + & \epsilon \left| \Omega (k,k+2)\right\rangle
\end{array}
\label{pba34}
\end{equation}
{\sf Now, we extend (\ref{pba25}), the definition of }$\Psi _{4\omega }${\sf %
\ , to } 
\begin{equation}
\Psi _{4\omega }(\xi _1,\xi _2)=\sum_{k_1<k_2}A(k_1,k_2)\left| \Omega
(k_1,k_2)\right\rangle  \label{pba35}
\end{equation}
{\sf Substituting (\ref{pba32}) and (\ref{pba34}) into the eigenvalue
equation, we obtain the following set of near equations } 
\begin{equation}
\left( E_{4\omega }-Q-Q^{-1}\right) A(k,k+1)=\epsilon A(k-1,k+1)+\epsilon
A(k,k+2)  \label{pba36}
\end{equation}
{\sf Using the same parametrization (\ref{pba28}) for these new
wavefunctions, the equation (\ref{pba36}) gives us the phase shift produced
by the interchange of the two interacting pseudoparticles } 
\begin{equation}
\frac{A_{21}}{A_{12}}=-\frac{\epsilon (1+\xi )+(Q+Q^{-1})\xi _2}{\epsilon
(1+\xi )+(Q+Q^{-1})\xi _1}  \label{pba37}
\end{equation}
{\sf We thus arrive to the Bethe Ansatz equations which fix the values of }$%
\xi _1${\sf \ and }$\xi _2${\sf \ in the energy equation (\ref{pba31}) } 
\begin{eqnarray}
\xi _2^N=-\frac{1+\xi +\epsilon (Q+Q^{-1})\xi _2}{1+\xi +\epsilon
(Q+Q^{-1})\xi _1}  \nonumber \\
\xi ^N=(\xi _1\xi _2)^N=1  \label{pba38}
\end{eqnarray}

\subsection{\sf Two-pseudoparticles and impurities}

{\sf In the sectors }$r>6\omega ${\sf , in addition the cases already
discussed, we find states with two interacting particles and impurities. Let
us now consider two pseudoparticles with one impurity of type }$\omega -1$%
{\sf . Theses eigenstates lie in the sector }$r=4\omega +1${\sf \ and we
seek them in the form } 
\begin{eqnarray}
\Psi _{4\omega +1}(\xi _1,\xi _2,\xi _3) &=&\sum_{k_1+1<k_2<k_3-2}A_{{\bf 1}%
}(k_1,k_2,k_3)\left| \Omega _1(k_1,k_2,k_3)\right\rangle  \nonumber \\
&&+\sum_{k_1+1<k_2<k_3}A_{{\bf 2}}(k_1,k_2,k_3)\left| \Omega
_2(k_1,k_2,k_3)\right\rangle  \nonumber \\
&&+\sum_{k_1+1<k_2<k_3-1}A_{{\bf 3}}(k_1,k_2,k_3)\left| \Omega
_3(k_1,k_2,k_3)\right\rangle  \label{pba39}
\end{eqnarray}
{\sf In }$A_{{\bf i}}(k_1,k_2,k_3)${\sf \ the index }${\bf i}=1,2,3${\sf \
characterizes the impurity position. Comparing (\ref{pba39}) with the state
build from the translational invariant products of two-pseudoparticles with
parameters }$\xi _2${\sf \ and }$\xi _3${\sf \ and one-impurity with
parameter }$\xi _1${\sf : } 
\begin{eqnarray}
\Psi _{4\omega +1}(\xi _1,\xi _2,\xi _3) &=&\left| (\omega -1)(\xi
_1)\right\rangle \times \Psi _{2\omega }(\xi _2)\times \Psi _{2\omega }(\xi
_3)  \nonumber \\
&&+\left| (\omega -1)(\xi _1)\right\rangle \times \Psi _{2\omega }(\xi
_3)\times \Psi _{2\omega }(\xi _2)  \nonumber \\
&&+\Psi _{2\omega }(\xi _2)\times \left| (\omega -1)(\xi _1)\right\rangle
\times \Psi _{2\omega }(\xi _3)  \nonumber \\
&&+\Psi _{2\omega }(\xi _3)\times \left| (\omega -1)(\xi _1)\right\rangle
\times \Psi _{2\omega }(\xi _2)  \nonumber \\
&&+\Psi _{2\omega }(\xi _2)\times \Psi _{2\omega }(\xi _3)\times \left|
(\omega -1)(\xi _1)\right\rangle  \nonumber \\
&&+\Psi _{2\omega }(\xi _3)\times \Psi _{2\omega }(\xi _2)\times \left|
(\omega -1)(\xi _1)\right\rangle  \label{pba40}
\end{eqnarray}

{\sf we get } 
\begin{eqnarray}
\left| \Omega _1(k_1,k_2,k_3)\right\rangle =\sum_{\alpha ,\ \beta =-\omega
}^\omega W(\alpha ,\beta ,\omega )\ \left| k_1(\omega -1),k_2(-\alpha
,\alpha ),k_3(-\beta ,\beta )\right\rangle  \nonumber \\
\left| \Omega _2(k_1,k_2,k_3)\right\rangle =\sum_{\alpha ,\ \beta =-\omega
}^\omega W(\alpha ,\beta ,\omega )\left| k_1(-\alpha ,\alpha ),k_2(\omega
-1),k_3(-\beta ,\beta )\right\rangle  \nonumber \\
\left| \Omega _3(k_1,k_2,k_3)\right\rangle =\sum_{\alpha ,\ \beta =-\omega
}^\omega W(\alpha ,\beta ,\omega )\left| k_1(-\alpha ,\alpha ),k_2(-\beta
,\beta ),k_3(\omega -1)\right\rangle  \label{pba41}
\end{eqnarray}
{\sf where } 
\begin{equation}
W(\alpha ,\beta ,\omega )=\epsilon (\alpha )\epsilon (\beta )\
q^{-<2e_\omega +e_\alpha +e_\beta ,\rho >}  \label{pba42}
\end{equation}
{\sf and the wavefunctions }$A_{{\bf i}}(k_1,k_2,k_3)${\sf \ which are
parametrized by plane waves as } 
\begin{eqnarray}
A_{{\bf 1}}(k_1,k_2,k_3)=A_{{\bf 1}23}\xi _1^{k_1}\xi _2^{k_2}\xi
_3^{k_3}+A_{{\bf 1}32}\xi _1^{k_1}\xi _2^{k_3}\xi _3^{k_2}  \nonumber \\
A_{{\bf 2}}(k_1,k_2,k_3)=A_{{\bf 2}13}\xi _1^{k_2}\xi _2^{k_1}\xi
_3^{k_3}+A_{{\bf 2}31}\xi _1^{k_2}\xi _2^{k_3}\xi _3^{k_1}  \nonumber \\
A_{{\bf 3}}(k_1,k_2,k_3)=A_{{\bf 3}12}\xi _1^{k_3}\xi _2^{k_1}\xi
_3^{k_2}+A_{{\bf 3}21}\xi _1^{k_3}\xi _2^{k_2}\xi _3^{k_1}.  \label{pba43}
\end{eqnarray}
{\sf Periodic boundary conditions read now } 
\begin{eqnarray}
A_{{\bf i}}(k_1,k_2,k_3)=A_{{\bf i}}(N+k_1,N+k_2,N+k_3),  \nonumber \\
A_{{\bf i}}(k_2,k_3,N+k_1)=A_{{\bf i}+{\bf 1}}(k_1,k_2,k_3),\quad \quad {\bf %
i}=1,2,3\ \bmod{3}{}  \label{pba44}
\end{eqnarray}
{\sf which imply that } 
\begin{eqnarray}
\xi _1^N=\frac{A_{{\bf 1}23}}{A_{{\bf 3}12}}=\frac{A_{{\bf 1}32}}{A_{{\bf 3}%
21}},\quad \xi _2^N=\frac{A_{{\bf 3}12}}{A_{{\bf 2}31}}=\frac{A_{{\bf 2}13}}{%
A_{{\bf 1}32}},\qquad \quad  \nonumber \\
\xi _3^N=\frac{A_{{\bf 3}21}}{A_{{\bf 2}13}}=\frac{A_{{\bf 2}31}}{A_{{\bf 1}%
23}},\quad \xi ^N=(\xi _1\xi _2\xi _3)^N=1  \label{pba45}
\end{eqnarray}

{\sf Action of }${\cal H}${\sf \ on the state }$\Psi _{4\omega +1}${\sf \
gives the following set of far equations: } 
\begin{eqnarray}
\left( E_{4\omega +1}-2Q-2Q^{-1}\right) A_1(k_1,k_2,k_3)=\epsilon
A_1(k_1,k_2-1,k_3)+\epsilon A_1(k_1,k_2+1,k_3)  \nonumber \\
+\epsilon A_1(k_1,k_2,k_3-1)+\epsilon A_1(k_1,k_2,k_3+1)  \nonumber \\
&&  \label{pba46}
\end{eqnarray}
{\sf and a similar set of eigenvalue equations for }$A_{{\bf 2}%
}(k_1,k_2,k_3) ${\sf \ and }$A_{{\bf 3}}(k_1,k_2,k_3)${\sf . The
parametrization (\ref{pba43}) solves these far equations provided that } 
\begin{equation}
E_{4\omega +1}=2Q+2Q^{-1}+\epsilon (\xi _2+\xi _2^{-1}+\xi _3+\xi _3^{-1})
\label{pba47}
\end{equation}
{\sf Taking into account the near equations we must split them in three
different neighborhood: (i) impurity neighbors of separated pseudoparticles
and, (ii) impurity far from neighbors pseudoparticles and (iii) when
impurity and pseudoparticles share the same neighborhood.}

{\sf In the case (i) we consider the second pseudoparticle far and follow
the steps for the case of one-pseudoparticle with impurity eigenstates.
Thus, the near equations can be read off from (\ref{pba36}) } 
\begin{eqnarray}
(E_{4\omega +1}-2Q-2Q^{-1})A_{{\bf 1}}(k,k+1,k_3)=\epsilon A_{{\bf 1}%
}(k,k+1,k_3-1)+\epsilon A_{{\bf 1}}(k,k+1,k_3+1)  \nonumber \\
+\epsilon A_{{\bf 1}}(k,k+2,k_3)+\epsilon A_{{\bf 2}}(k,k+2,k_3)
\label{pba48}
\end{eqnarray}
{\sf and a similar set of equations coupling }$A_{{\bf 2}}${\sf \ and }$A_{%
{\bf 3}}${\sf . It follows from the consistency between (\ref{pba46}) and (%
\ref{pba48}) that } 
\begin{equation}
A_{{\bf 1}}(k,k,k_3)\equiv A_{{\bf 2}}(k,k+2,k_3)  \label{pba49}
\end{equation}
{\sf and similar identification between }$A_{{\bf 2}}${\sf \ and }$A_{{\bf 3}%
}${\sf . The plane waves (\ref{pba43}) solve these identifications provided }
\begin{equation}
\xi _1^2=\frac{A_{{\bf 1}23}}{A_{{\bf 2}13}}=\frac{A_{{\bf 1}32}}{A_{{\bf 2}%
31}}.=\frac{A_{{\bf 2}31}}{A_{{\bf 3}21}}=\frac{A_{{\bf 2}13}}{A_{{\bf 3}12}}
\label{pba50}
\end{equation}

{\sf For the case (ii) we can derive the near equations from those of
two-pseudoparticles case. Keeping the impurity far and following the steps (%
\ref{pba30})--(\ref{pba37}) we get } 
\begin{equation}
\left( E_{4\omega +1}-Q-Q^{-1}\right) A_1(k_1,k,k+1)=\epsilon
A_1(k_1,k-1,k+1)+\epsilon A_1(k_1,k,k+2)  \label{pba51}
\end{equation}
{\sf and a similar set of equations for }$A_{{\bf 2}}${\sf \ and }$A_{{\bf 3}%
}${\sf . The case (iii) is obtained from (\ref{pba51}) for }$k_1=k-1${\sf .}

{\sf The parametrization (\ref{pba43}) solves this provided that } 
\begin{equation}
\frac{A_{{\bf 1}23}}{A_{{\bf 1}32}}=-\ \frac{\epsilon (1+\xi _3\xi
_2)+(Q+Q^{-1})\xi _2}{\epsilon (1+\xi _2\xi _3)+(Q+Q^{-1})\xi _3}
\label{pba52}
\end{equation}
{\sf Matching the constraint equations (\ref{pba52}), (\ref{pba50}) and (\ref
{pba45}) we arrive to the Bethe equations } 
\begin{eqnarray}
\xi _a^N\xi _1^2=-\ \frac{1+\xi _b\xi _a+\epsilon (Q+Q^{-1})\xi _a}{1+\xi
_a\xi _b+\epsilon (Q+Q^{-1})\xi _b},\quad a\neq b=2,3  \label{pba53} \\
\xi ^N=(\xi _1\xi _2\xi _3)^N=1,\qquad \xi _1^{N-4}=1.
\end{eqnarray}
{\sf The origin of the exponent (}$N-4${\sf ) in the impurity parameter can
be understood by saying that after the two pseudoparticles propagate past
the impurity, the position of impurity is shifted by four lattice sites.}

{\sf Next, we can also find eigenstates with two pseudoparticles and more
than one impurities. They can be described in the following way: Let us
consider an eigenstate with }$l>1${\sf \ impurities with parameters }$\xi
_1,\xi _2,\cdots ,\xi _l${\sf \ and two pseudoparticles with parameters }$%
\xi _{l+1}${\sf \ and }$\xi _{l+2}${\sf . The energy eigenvalue is } 
\begin{equation}
E_r=2Q+2Q^{-1}+\epsilon \left( \xi _{l+1}+\xi _{l+1}^{-1}+\xi _{l+2}+\xi
_{l+2}^{-1}\right)  \label{pba54}
\end{equation}
{\sf and the Bethe equations } 
\begin{eqnarray}
\xi _{l+1}^N\xi _1^2\xi _2^2\cdots \xi _l^2=-\ \frac{1+\xi _{l+1}\xi
_{l+2}+\epsilon (Q+Q^{-1})\xi _{l+1}}{1+\xi _{l+1}\xi _{l+2}+\epsilon
(Q+Q^{-1})\xi _{l+2}}  \nonumber \\
\xi _a^{N-4}=1,\quad a=1,2,...,l  \label{pba55}
\end{eqnarray}
{\sf Moreover, }$\xi ^N=1${\sf \ with }$\ \xi =\xi _1\xi _2\cdots \xi _{l+2}$%
{\sf .}

\subsection{\sf Three-pseudoparticle eigenstates}

{\sf In the sector }$r=6\omega ${\sf , in addition to the previously
discussed eigenstates of one and two pseudoparticles with impurities, one
can find eigenstates with three interacting pseudoparticles with parameters }%
$\xi _1,\xi _2${\sf \ and }$\xi _3${\sf . We start seek them in the form } 
\begin{equation}
\Psi _{6\omega }(\xi _1,\xi _2,\xi _3)=\sum_{k_1+2\leq k_2\leq
k_3-2}A(k_1,k_2,k_3)\left| \Omega (k_1,k_2,k_3)\right\rangle  \label{3p}
\end{equation}
{\sf where }$\left| \Omega (k_1,k_2,k_3)\right\rangle =\otimes
_{i=1}^3\left| \Omega (k_i)\right\rangle ${\sf \ . The corresponding
wavefunctions } 
\begin{equation}
\begin{array}{lll}
A(k_1,k_2,k_3) & = & A_{123}\xi _1^{k_1}\xi _2^{k_2}\xi _3^{k_3}+A_{132}\xi
_1^{k_1}\xi _2^{k_3}\xi _3^{k_2}+A_{213}\xi _1^{k_2}\xi _2^{k_1}\xi _3^{k_3}
\\ 
&  &  \\ 
&  & +A_{231}\xi _1^{k_2}\xi _2^{k_3}\xi _3^{k_1}+A_{312}\xi _1^{k_3}\xi
_2^{k_1}\xi _3^{k_2}+A_{321}\xi _1^{k_3}\xi _2^{k_2}\xi _3^{k_1}
\end{array}
\label{3pw}
\end{equation}
{\sf satisfy the periodic boundary conditions } 
\begin{equation}
A(k_2,k_3,N+k_1)=A(k_1,k_2,k_3),\ A(N+k_1,N+k_2,N+k_3)=A(k_1,k_2,k_3)
\end{equation}
{\sf which imply that } 
\begin{eqnarray}
\xi _1^N &=&\frac{A_{123}}{A_{312}}=\frac{A_{132}}{A_{321}},\quad \xi _2^N=%
\frac{A_{312}}{A_{231}}=\frac{A_{213}}{A_{132}},\qquad  \nonumber \\
&&  \nonumber \\
\xi _3^N &=&\frac{A_{321}}{A_{213}}=\frac{A_{231}}{A_{123}},\quad \xi
^N=(\xi _1\xi _2\xi _3)^N=1  \label{3pbc}
\end{eqnarray}
{\sf These relations show us that the interchange of two-pseudoparticles is
independent of the position of the third particle.}

{\sf Applying }${\cal H}${\sf \ to (\ref{3p}), we obtain a set of equations
for }$A(k_1,k_2,k_3)${\sf . When the three pseudoparticles are separated, }$%
(k_1+2<k_2<k_3-2)${\sf , we get the following far equations: } 
\begin{eqnarray}
(E_{6s}-3Q-3Q^{-1})A(k_1,k_2,k_3) &=&\epsilon A(k_1-1,k_2,k_3)+\epsilon
A(k_1+1,k_2,k_3)  \nonumber \\
&&+\epsilon A(k_1,k_2-1,k_3)+\epsilon A(k_1,k_2+1,k_3)  \nonumber \\
&&+\epsilon A(k_1,k_2,k_3-1)+\epsilon A(k_1,k_2,k_3+1)  \nonumber \\
&&  \label{3pe}
\end{eqnarray}
{\sf It is simple verify that the wavefunctions (\ref{3pw}) satisfy these
far equations provided } 
\begin{equation}
E_{6\omega }=\sum_{n=1}^3\left\{ Q+Q^{-1}+\xi _n+\xi _n^{-1}\right\}
\label{3eq}
\end{equation}

{\sf Applying }${\cal H}${\sf \ on the near states we get the following set
equations: } 
\begin{eqnarray}
{\cal H}\left| \Omega (k_1,k_1+2,k_3)\right\rangle &=&(2Q+2Q^{-1})\left|
\Omega (k_1,k_1+2,k_3)\right\rangle +\epsilon \left| \Omega
(k_1-1,k_1+2,k_3)\right\rangle  \nonumber \\
&&+\epsilon \left| \Omega (k_1,k_1+3,k_3)\right\rangle +\epsilon \left|
\Omega (k_1,k_1+2,k_3-1)\right\rangle  \nonumber \\
&&+\epsilon \left| \Omega (k_1,k_1+2,k_3+1)\right\rangle +U_{k_1+1}\left|
\Omega (k_1,k_1+2,k_3)\right\rangle  \nonumber \\
&&
\end{eqnarray}
{\sf for }$k_3>k_1+4${\sf , which correspond to the meeting of two
pseudoparticles at the left of the third pseudoparticle, which is far from
of the meeting position. } 
\begin{eqnarray}
{\cal H}\left| \Omega (k_1,k_2,k_2+2)\right\rangle &=&(2Q+2Q^{-1})\left|
\Omega (k_1,k_2,k_2+2)\right\rangle +\epsilon \left| \Omega
(k_1-1,k_2,k_2+2)\right\rangle  \nonumber \\
&&+\epsilon \left| \Omega (k_1+1,k_2,k_2+2)\right\rangle +\epsilon \left|
\Omega (k_1,k_2-1,k_2+2)\right\rangle  \nonumber \\
&&+\epsilon \left| \Omega (k_1,k_2,k_2+3)\right\rangle +U_{k_2+1}\left|
\Omega (k_1,k_2,k_2+2)\right\rangle  \nonumber \\
&&
\end{eqnarray}
{\sf for }$k_2>k_1+2${\sf , which correspond to the meeting of two
pseudoparticles at the right of the far pseudoparticle. Moreover, there is
one set of equations which correspond to the meeting of three
pseudoparticles } 
\begin{eqnarray}
{\cal H}\left| \Omega (k,k+2,k+4)\right\rangle &=&(Q+Q^{-1})\left| \Omega
(k,k+2,k+4)\right\rangle +\epsilon \left| \Omega (k-1,k+2,k+4)\right\rangle 
\nonumber \\
&&+\epsilon \left| \Omega (k,k+2,k+5)\right\rangle +U_{k+1}\left| \Omega
(k,k+2,k+4)\right\rangle  \nonumber \\
&&+U_{k+3}\left| \Omega (k,k+2,k+4)\right\rangle
\end{eqnarray}
{\sf In deriving these equations new states made their debut. In order to
incorporate these new states in the eigenvalue problem we define: } 
\begin{eqnarray}
U_{k_1+1}\left| \Omega (k_1,k_1+2,k_3)\right\rangle &=&\epsilon \left|
\Omega (k_1,k_1+1,k_3)\right\rangle +\epsilon \left| \Omega
(k_1+1,k_1+2,k_3)\right\rangle  \nonumber \\
U_{k_2+1}\left| \Omega (k_1,k_2,k_2+2)\right\rangle &=&\epsilon \left|
\Omega (k_1,k_2+1,k_2+2)\right\rangle +\epsilon \left| \Omega
(k_1,k_2,k_2+1)\right\rangle  \nonumber \\
U_{k+1}\left| \Omega (k,k+2,k+4)\right\rangle &=&\epsilon \left| \Omega
(k,k+1,k+4)\right\rangle +\epsilon \left| \Omega (k+1,k+2,k+4)\right\rangle 
\nonumber \\
U_{k+3}\left| \Omega (k,k+2,k+4)\right\rangle &=&\epsilon \left| \Omega
(k,k+3,k+4)\right\rangle +\epsilon \left| \Omega (k,k+2,k+3)\right\rangle
\end{eqnarray}

{\sf Applying }${\cal H}${\sf \ to these new states the result can be
incorporated to the eigenvalue problem provided the definition of }$\Psi
_{6\omega }${\sf \ (\ref{3p}) is extended to } 
\begin{equation}
\Psi _{6\omega }(\xi _1,\xi _2,\xi
_3)=\sum_{k_1<k_2<k_3}A(k_1,k_2,k_3)\left| \Omega (k_1,k_2,k_3)\right\rangle
\end{equation}
{\sf After this we are left with three meeting equations } 
\begin{eqnarray}
&&\left. (E_{6\omega }-2Q-2Q^{-1})A(k_1,k_1+1,k_3)=\epsilon
A(k_1-1,k_1+1,k_3)+\epsilon A(k_1,k_1+2,k_3)\right.  \nonumber \\
&&\hspace{5cm}{}+\epsilon A(k_1,k_1+1,k_3-1)+\epsilon A(k_1,k_1+1,k_3+1) 
\nonumber \\
&&
\end{eqnarray}
{\sf for }$k_3>k_1+2${\sf , } 
\begin{eqnarray}
&&\left. (E_{6\omega }-2Q-2Q^{-1})A(k_1,k_2,k_2+1)=\epsilon
A(k_1,k_2-1,k_2+1)+\epsilon A(k_1,k_2,k_2+2)\right.  \nonumber \\
&&\hspace{5cm}{}+\epsilon A(k_1-1,k_2,k_2+1)+\epsilon A(k_1+1,k_2,k_2+1) 
\nonumber \\
&&
\end{eqnarray}
{\sf for }$k_1+2<k_2${\sf \ and } 
\[
(E_{6\omega }-Q-Q^{-1})A(k,k+1,k+2)=\epsilon A(k-1,k+1,k+2)+\epsilon
A(k,k+1,k+3) 
\]

{\sf It is easy to verify that the parametrization (\ref{3pw}) and (\ref{3eq}%
) solve these equations provided } 
\begin{eqnarray}
\frac{A_{123}}{A_{213}} &=&\frac{A_{231}}{A_{321}}=-\frac{1+\xi _1\xi
_2+\epsilon (Q+Q^{-1})\xi _1}{1+\xi _1\xi _2+\epsilon (Q+Q^{-1})\xi _2} 
\nonumber \\
&&  \nonumber \\
\frac{A_{132}}{A_{231}} &=&\frac{A_{213}}{A_{312}}=-\frac{1+\xi _1\xi
_3+\epsilon (Q+Q^{-1})\xi _1}{1+\xi _1\xi _3+\epsilon (Q+Q^{-1})\xi _3} 
\nonumber \\
&&  \nonumber \\
\frac{A_{312}}{A_{321}} &=&\frac{A_{123}}{A_{132}}=-\frac{1+\xi _2\xi
_3+\epsilon (Q+Q^{-1})\xi _2}{1+\xi _2\xi _3+\epsilon (Q+Q^{-1})\xi _3}
\end{eqnarray}
{\sf Matching these constraints and the periodic boundary conditions (\ref
{3pbc}) we get the Bethe Ansatz equations } 
\begin{eqnarray}
\xi _a^N &=&\prod_{b\neq a=1}^3\left\{ -\frac{1+\xi _a\xi _b+\epsilon
(Q+Q^{-1})\xi _a}{1+\xi _a\xi _b+\epsilon (Q+Q^{-1})\xi _b}\right\} ,\quad
a=1,2,3  \nonumber \\
(\xi _1\xi _2\xi _3)^N &=&1
\end{eqnarray}

\subsection{\sf General eigenstates}

{\sf The generalization to any }$r${\sf \ is now immediate. Since the
Yang-Baxter equations are satisfied, there is only two-pseudoparticle
scattering (using the }$S${\sf -matrix language). Therefore, neighbor
equations, where more then two pseudoparticles become neighbors, are nor
expected to give any new restrictions. For instance, in the sector }$%
r=6\omega ${\sf , we saw that the interchange of two-pseudoparticles is
independent of the position of the third particle. Thus, in a sector with }$%
p ${\sf \ pseudoparticles we expect that the }$p${\sf -pseudoparticle phase
shift will be a sum of }$p(p-1)/2${\sf \ two-pseudoparticle phase shift. The
energy is given by the sum of single pseudoparticle energies. The
corresponding Bethe Ansatz equations depend on the phase shift of two
pseudoparticles and on the number of impurity. For a generic sector one can
verify that no different neighborhood those discussed above can appear. So,
no additional meeting conditions will be encountered. Thus, we can extend
the previous results to the }$p${\sf \ -pseudoparticle states in the
following way: In a generic sector }$r${\sf \ with }$l${\sf \ impurities
parametrized by }$\xi _1\xi _2\cdots \xi _l${\sf \ and }$p${\sf \
pseudoparticles with parameters }$\xi _{l+1}\xi _{l+2}\cdots \xi _{l+p}${\sf %
, the energy is } 
\begin{equation}
E_r=\sum_{n=l+1}^p\left\{ Q+Q^{-1}+\epsilon \left( \xi _n+\xi _n^{-1}\right)
\right\}
\end{equation}
{\sf with }$\xi _n${\sf \ determined by the Bethe ansatz equations } 
\begin{eqnarray}
\xi _a^N\xi _1^2\xi _2^2\cdots \xi _l^2 &=&\prod_{b\neq a=l+1}^{l+p}\left\{ -%
\frac{1+\xi _b\xi _a+\epsilon (Q+Q^{-1})\xi _a}{1+\xi _a\xi _b+\epsilon
(Q+Q^{-1})\xi _b}\right\}  \nonumber \\
\xi _c^{N-2p} &=&1,\quad c=1,2,...,l  \nonumber \\
\xi ^N &=&1,\qquad \xi =\xi _1\xi _2\cdots \xi _l\xi _{l+1}\xi _{l+2}\cdots
\xi _{l+p}.  \label{pba60}
\end{eqnarray}
{\sf The energy eigenvalues and the Bethe equations depend on the
deformation parameter }$q${\sf , through the relation (\ref{int3}): } 
\begin{equation}
Q+Q^{-1}=\left\{ 
\begin{array}{lll}
\lbrack 2s+1] & {\rm for} & A_1 \\ 
\lbrack 2n-1][n+1/2]/[n-1/2] & {\rm for} & B_n(n\geq 2) \\ 
\lbrack n][2n+2]/[n+1] & {\rm for} & C_n(n\geq 1) \\ 
\lbrack n][2n-2]/[n-1] & {\rm for} & D_n(n\geq 3)
\end{array}
\right.  \label{pba61}
\end{equation}

{\sf We obtained thus the spectra with periodic boundary conditions of
quantum spin-chain models, arising as representations of the Temperley-Lieb
algebra. As expected, all these models have equivalent spectra up to
degeneracies of their eigenvalues. From a suitable sorting of the parameters 
}$\xi _i${\sf , one can insure that the spectra of lower-}$r${\sf \ sectors
are contained entirely in the higher-}$r${\sf \ sectors.}

\section{\sf Bethe Ansatz: Non-local boundary conditions}

{\sf It is the purpose of this section to present and solve, via coordinate
Bethe ansatz, the quantum group invariant closed TL Hamiltonians which can
be written as \cite{Martin}:}

\begin{equation}
{\cal H}=\sum_{k=1}^{N-1}U_k+{\cal U}_0  \label{cbah}
\end{equation}
{\sf where }$U_k${\sf \ is a Temperley-Lieb operator and }${\bf bt}={\cal U}%
_0${\sf \ is non-local term defined through of a operator }$G${\sf \ which
plays the role of the translation operator } 
\begin{equation}
{\cal U}_0=GU_{N-1}G^{-1}\quad ,\quad G=(Q-U_1)(Q-U_2)\cdots (Q-U_{N-1})
\label{cba1}
\end{equation}
{\sf satisfying }$[{\cal H},G]=0${\sf \ and additionally invariance with
respect to the quantum algebra. The operator }$G${\sf \ shifts the }$U_k$%
{\sf \ by one unit }$GU_kG^{-1}=U_{k+1}${\sf \ and maps }${\cal U}_0${\sf \
into }$U_1${\sf , which manifest the translational invariance of }${\cal H}$%
{\sf . In this sense the Hamiltonian (\ref{cbah}) is periodic.}

{\sf For the }$q${\sf -deformed }$A${\sf -}$D${\sf \ Temperley-Lieb algebra,
the matrix elements of }$U_k${\sf \ is again given by (\ref{int9}), i.e. : } 
\begin{eqnarray}
\langle \nu ,\lambda |U|\mu ,\kappa \rangle =\epsilon (\mu )\epsilon (\nu
)q^{-<e_\mu +e_\nu ,\rho >}\delta _{\mu +\kappa ,0}\,\delta _{\nu +\lambda
,0}  \nonumber \\
\mu ,\nu ,\kappa ,\lambda \in J  \label{cba2}
\end{eqnarray}
{\sf From (\ref{cba2}) we choose an particular Bethe state } 
\begin{equation}
\left| \Omega (k)\right\rangle =\sum_{\alpha =-\omega }^\omega \epsilon
(\omega )\epsilon (\alpha )\ q^{-<e_\omega +e_\alpha ,\rho >}\left|
k(-\alpha ,\alpha )\right\rangle  \label{cba3}
\end{equation}
{\sf which is an eigenstate of }$U_k${\sf \ and it is shifted by one unit
under the action of }$U_{k\pm 1}$

\begin{eqnarray}
U_k\left| \Omega (k)\right\rangle =(Q+Q^{-1})\left| \Omega (k)\right\rangle 
\nonumber \\
U_{k\pm 1}\left| \Omega (k)\right\rangle =\epsilon \left| \Omega (k\pm
1)\right\rangle \quad ,\quad U_k\left| \Omega (k\pm )\right\rangle =\epsilon
\left| \Omega (k)\right\rangle  \nonumber \\
U_k\left| \Omega (l)\right\rangle =0\quad {for\quad }k\neq \{l-1,\ l,\ l+1\}
\label{cba4}
\end{eqnarray}
{\sf where }$\epsilon =1${\sf \ for }$B_n${\sf , }$D_n${\sf \ and }$A_1${\sf %
\ ( }$s${\sf \ integer) and }$\epsilon =-1${\sf \ for }$C_n${\sf \ and }$A_1$%
{\sf \ ( }$s${\sf \ semi-integer).}

{\sf The action of the operator }$G${\sf \ on the states }$\left| \Omega
(k)\right\rangle ${\sf \ can be easily computed using (\ref{cba4}): It is
simple on the bulk and at the left boundary}

\begin{equation}
G\left| \Omega (k)\right\rangle =-\epsilon Q^{N-2}\ \left| \Omega
(k+1)\right\rangle {\rm \quad },{\rm \quad }1\leq k\leq N-2  \label{cba5}
\end{equation}
{\sf but manifests its nonlocality at the right boundary}

\begin{equation}
G\left| \Omega (N-1)\right\rangle =\epsilon
Q^{N-2}\sum_{k=1}^{N-1}(-\epsilon Q)^{-k}\ \left| \Omega (N-k)\right\rangle
\label{cba6}
\end{equation}
{\sf Similarly, the action of the operator }$G^{-1}=(Q^{-1}-U_{N-1})\cdots
(Q^{-1}-U_1)${\sf \ is simple on the bulk and at the right boundary}

\begin{equation}
G^{-1}\left| \Omega (k)\right\rangle =-\epsilon Q^{-N+2}\left| \Omega
(k-1)\right\rangle \ \quad ,\quad 2\leq k\leq N-1  \label{cba7}
\end{equation}
{\sf and non-local at the left boundary } 
\begin{equation}
G^{-1}\left| \Omega (1)\right\rangle =\epsilon
Q^{-N+2}\sum_{k=1}^{N-1}(-\epsilon Q)^k\ \left| \Omega (k)\right\rangle .
\label{cba8}
\end{equation}

{\sf Now we proceed the diagonalization of }${\cal H}${\sf \ as was made for
the periodic case. As (\ref{cbah}) and (\ref{pbah}) have the same bulk,\
i.e., differences came from the boundary terms, we will keep all results
relating to the bulk of the periodic case presented in the previous section.}

\subsection{\sf One-pseudoparticle eigenstates}

{\sf Let us consider one free pseudoparticle which lies in the sector }$%
r=2\omega $

\begin{equation}
\Psi _{2\omega }=\sum_{k=1}^{N-1}A(k)\left| \Omega (k)\right\rangle .
\label{cba9}
\end{equation}
{\sf The action of the operator }${\cal U}=\sum_{k=1}^{N-1}U_k${\sf \ on the
states }$\left| \Omega (k)\right\rangle ${\sf \ can be computed using (\ref
{cba4}):}

\begin{eqnarray}
{\cal U}\left| \Omega (1)\right\rangle =(Q+Q^{-1})\left| \Omega
(1)\right\rangle +\epsilon \left| \Omega (2)\right\rangle  \nonumber \\
{\cal U}\left| \Omega (k)\right\rangle =(Q+Q^{-1})\left| \Omega
(k)\right\rangle +\epsilon \left| \Omega (k-1)\right\rangle +\epsilon \left|
\Omega (k+1)\right\rangle  \nonumber \\
\qquad \quad {{\rm for}\ }2\leq k\leq N-2  \nonumber \\
{\cal U}\left| \Omega (N-1)\right\rangle =(Q+Q^{-1})\left| \Omega
(N-1)\right\rangle +\epsilon \left| \Omega (N-2)\right\rangle .
\label{cba10}
\end{eqnarray}
{\sf and using (\ref{cba5})--(\ref{cba8}) one can see that the action of }$%
{\cal U}_0=GU_{N-1}G^{-1}${\sf \ vanishes on the bulk } 
\begin{equation}
{\cal U}_0\left| \Omega (k)\right\rangle =0\quad ,\quad 2\leq k\leq N-2
\label{cba13}
\end{equation}
{\sf and is nonlocal at the boundaries } 
\begin{equation}
{\cal U}_0\left| \Omega (1)\right\rangle =-\epsilon \sum_{k=1}^{N-1}\
(-\epsilon Q)^k\ \left| \Omega (k)\right\rangle ,\quad {\cal U}_0\left|
\Omega (N-1)\right\rangle =-\epsilon \sum_{k=1}^{N-1}\ (-\epsilon Q)^{-N+k}\
\left| \Omega (k)\right\rangle .  \label{cba14}
\end{equation}
{\sf which are connected by } 
\begin{equation}
{\cal U}_0\left| \Omega (N-1)\right\rangle =(-\epsilon Q)^{-N}\ {\cal U}%
_0\left| \Omega (1)\right\rangle .  \label{cba15}
\end{equation}
{\sf From these equations we can understood the role of }${\cal U}_0${\sf :
Although the Hamiltonian (\ref{cbah}) is a global operator, it manifests the
property of essential locality. From the physical point of view, this type
of models exhibit behavior similar to closed chains with twisted boundary
conditions.}

{\sf Before we substitute these results into the eigenvalue equation, we
will define two new states}

\begin{equation}
\epsilon \left| \Omega (0)\right\rangle ={\cal U}_0\left| \Omega
(1)\right\rangle ,\quad \ \epsilon \left| \Omega (N)\right\rangle ={\cal U}%
_0\left| \Omega (N-1)\right\rangle  \label{cba16}
\end{equation}
{\sf to include the cases }$k=0${\sf \ and }$k=N${\sf \ into the definition
of }$\Psi _{2\omega }${\sf , equation (\ref{cba9}). Finally, the action of }$%
{\cal H}={\cal U}+{\cal U}_0${\sf \ on the states }$\left| \Omega
(k)\right\rangle ${\sf \ is}

\begin{eqnarray}
{\cal H}\left| \Omega (0)\right\rangle =(Q+Q^{-1})\left| \Omega
(0)\right\rangle +(-\epsilon Q)^N\epsilon \left| \Omega (N-1)\right\rangle
+\epsilon \left| \Omega (1)\right\rangle  \nonumber \\
&&  \nonumber \\
{\cal H}\left| \Omega (k)\right\rangle =(Q+Q^{-1})\left| \Omega
(k)\right\rangle +\epsilon \left| \Omega (k-1)\right\rangle +\epsilon \left|
\Omega (k+1)\right\rangle  \nonumber \\
\qquad \qquad {\rm for\ }1\leq k\leq N-2  \nonumber \\
&&  \nonumber \\
{\cal H}\left| \Omega (N-1)\right\rangle =(Q+Q^{-1})\left| \Omega
(N-1)\right\rangle +\epsilon \left| \Omega (N-2)\right\rangle +(-\epsilon
Q)^{-N}\epsilon \left| \Omega (0)\right\rangle  \nonumber \\
&&  \nonumber \\
{\cal H}\left| \Omega (N)\right\rangle =(Q+Q^{-1})\left| \Omega
(N)\right\rangle +\epsilon \left| \Omega (N-1)\right\rangle +(-\epsilon
Q)^{-N}\epsilon \left| \Omega (1)\right\rangle  \label{cba17}
\end{eqnarray}
{\sf Substituting these results into the eigenvalue equation\ }${\cal H}\Psi
_{2\omega }=E_{2\omega }\ \Psi _{2\omega }${\sf \ we get a complete set of
eigenvalue equations for the wavefunctions}

\begin{eqnarray}
E_{2s}\ A(k)=(Q+Q^{-1})A(k)+\epsilon A(k-1)+\epsilon A(k+1)  \nonumber \\
\quad \qquad {\rm for\ }1\leq k\leq N-1  \label{cba18}
\end{eqnarray}
{\sf provided the following boundary conditions } 
\begin{equation}
(-\epsilon Q)^NA(k)=A(N+k)  \label{cba20}
\end{equation}
{\sf are satisfied.}

{\sf The plane wave parametrization }$A(k)=A\xi ^k${\sf \ solves these
eigenvalue equations and the boundary conditions provided that: } 
\begin{eqnarray}
E_{2\omega }=Q+Q^{-1}+\epsilon (\xi +\xi ^{-1})\quad  \nonumber \\
\xi ^N=(-\epsilon Q)^N  \label{cba21}
\end{eqnarray}
{\sf where }$\xi ={\rm e}^{i\theta }${\sf \ and }$\theta ${\sf \ being the
momentum.}

\subsection{\sf Two-pseudoparticle eigenstates}

{\sf Let us now consider the sector }$r=4\omega ${\sf , where we can find an
eigenstate with two interacting pseudoparticles. We seek the corresponding
eigenfunction as products of single pseudoparticles eigenfunctions, i.e. } 
\begin{equation}
\Psi _{4\omega }=\sum_{k_1+1<k_2}A(k_1,k_2)\left| \Omega
(k_1,k_2)\right\rangle  \label{cba22}
\end{equation}
{\sf where } 
\begin{equation}
\left| \Omega (k_1,k_2)\right\rangle =\sum_{\alpha ,\ \beta =-\omega
}^\omega \epsilon (\alpha )\epsilon (\beta )\ q^{-<2e_\omega +e_\alpha
+e_\beta ,\rho >}\left| k_1(-\alpha ,\alpha ),k_2(-\beta ,\beta
)\right\rangle  \label{cba23}
\end{equation}

{\sf To solve the eigenvalue equation }${\cal H}\Psi _{4\omega }=E_{4\omega
}\Psi _{4\omega }${\sf , we recall (\ref{cba4}) to get the action of }${\cal %
U}${\sf \ and }${\cal U}_0${\sf \ on the states }$\left| \Omega
(k_1,k_2)\right\rangle ${\sf . Here we have to consider four cases: (i)\
when the two pseudoparticles are separated in the bulk, the action of }$%
{\cal U}${\sf \ is } 
\begin{eqnarray}
{\cal U}\left| \Omega (k_1,k_2)\right\rangle =2(Q+Q^{-1})\left| \Omega
(k_1,k_2)\right\rangle +\epsilon \left| \Omega (k_1-1,k_2)\right\rangle
+\epsilon \left| \Omega (k_1+1,k_2)\right\rangle  \nonumber \\
+\epsilon \left| \Omega (k_1,k_2-1)\right\rangle +\epsilon \left| \Omega
(k_1,k_2+1)\right\rangle  \label{cba24}
\end{eqnarray}
{\sf i.e., for }$k_1${\sf \ }$\geq 2${\sf \ and }$k_1+3\leq k_2\leq N-2${\sf %
; (ii) when the two pseudoparticles are separated but one of them or both
are at the boundaries } 
\begin{eqnarray}
{\cal U}\left| \Omega (1,k_2)\right\rangle =2(Q+Q^{-1})\left| \Omega
(1,k_2)\right\rangle +\epsilon \left| \Omega (2,k_2)\right\rangle +\epsilon
\left| \Omega (1,k_2-1)\right\rangle  \nonumber \\
+\epsilon \left| \Omega (1,k_2+1)\right\rangle  \label{cba25}
\end{eqnarray}
\begin{eqnarray}
{\cal U}\left| \Omega (k_1,N-1)\right\rangle =2(Q+Q^{-1})\left| \Omega
(k_1,N-1)\right\rangle +\epsilon \left| \Omega (k_1-1,N-1)\right\rangle 
\nonumber \\
+\epsilon \left| \Omega (k_1+1,N-1)\right\rangle +\epsilon \left| \Omega
(k_1,N-2)\right\rangle  \label{cba26}
\end{eqnarray}
\begin{equation}
{\cal U}\left| \Omega (1,N-1)\right\rangle =2(Q+Q^{-1})\left| \Omega
(1,N-1)\right\rangle +\epsilon \left| \Omega (2,N-1)\right\rangle +\epsilon
\left| \Omega (1,N-2)\right\rangle  \label{ba27}
\end{equation}
{\sf where }$2\leq k_1\leq N-4${\sf \ and }$4\leq k_2\leq N-2${\sf ; (iii)
when the two pseudoparticles are neighbors in the bulk } 
\begin{eqnarray}
{\cal U}\left| \Omega (k,k+2)\right\rangle =2(Q+Q^{-1})\left| \Omega
(k,k+2)\right\rangle +\epsilon \left| \Omega (k-1,k+2)\right\rangle
+\epsilon \left| \Omega (k,k+3)\right\rangle  \nonumber \\
+U_{k+1}\left| \Omega (k,k+2)\right\rangle  \label{cba28}
\end{eqnarray}
{\sf for }$2\leq k\leq N-4${\sf \ and (iv) when the two pseudoparticles are
neighbors and at the boundaries } 
\begin{equation}
{\cal U}\left| \Omega (1,3)\right\rangle =2(Q+Q^{-1})\left| \Omega
(1,3)\right\rangle +\epsilon \left| \Omega (1,4)\right\rangle +U_2\left|
\Omega (1,3)\right\rangle  \label{cba29}
\end{equation}
\begin{eqnarray}
{\cal U}\left| \Omega (N-3,N-1)\right\rangle =2(Q+Q^{-1})\left| \Omega
(N-3,N-1)\right\rangle +\epsilon \left| \Omega (N-4,N-1)\right\rangle 
\nonumber \\
+U_{N-2}\left| \Omega (N-3,N-1)\right\rangle  \label{cba30}
\end{eqnarray}

{\sf Moreover, the action of }${\cal U}_0${\sf \ does not depend on the
pseudoparticles are neither separated nor neighbors. It is vanishes in the
bulk } 
\begin{equation}
{\cal U}_0\left| \Omega (k_1,k_2)\right\rangle =0\quad {{\rm for}\quad }%
k_1\neq 1\ {\rm and}k_2\neq N-1,  \label{cba31}
\end{equation}
{\sf and different of zero at the boundaries: } 
\begin{eqnarray}
{\cal U}_0\left| \Omega (1,k_2)\right\rangle &=&-\epsilon
\sum_{k=1}^{k_2-2}(-\epsilon Q)^k\left| \Omega (k,k_2)\right\rangle
-(-\epsilon Q)^{k_2-1}U_{k_2}\left| \Omega (k_2-1,k_2+1)\right\rangle 
\nonumber \\
&&-\epsilon \sum_{k=k_2+2}^{N-1}(-\epsilon Q)^{k-2}\left| \Omega
(k_2,k)\right\rangle  \label{cba32}
\end{eqnarray}
\begin{equation}
{\cal U}_0\left| \Omega (k_1,N-1)\right\rangle =(-\epsilon Q)^{-N+2}\ {\cal U%
}_0\left| \Omega (1,k_2)\right\rangle  \label{cba33}
\end{equation}
{\sf where }$2\leq k_1\leq N-3${\sf \ and }$3\leq k_2\leq N-2${\sf .}

{\sf Following the same procedure of one-pseudoparticle case we again define
new states in order to have consistency between bulk and boundaries terms } 
\begin{eqnarray}
{\cal U}_0\left| \Omega (1,k_2)\right\rangle &=&\epsilon \left| \Omega
(0,k_2)\right\rangle ,\quad {\cal U}_0\left| \Omega (k_1,N-1)\right\rangle
=\epsilon \left| \Omega (k_1,N)\right\rangle  \nonumber \\
{\cal U}_0\left| \Omega (1,N-1)\right\rangle &=&\epsilon \left| \Omega
(0,N-1)\right\rangle +\epsilon \left| \Omega (1,N)\right\rangle  \nonumber \\
U_{k+1}\left| \Omega (k,k+2)\right\rangle &=&\epsilon \left| \Omega
(k,k+1)\right\rangle +\epsilon \left| \Omega (k+1,k+2)\right\rangle
\label{cba34}
\end{eqnarray}
{\sf Acting with }${\cal H}${\sf \ on these new states, we get } 
\begin{eqnarray}
{\cal H}\left| \Omega (0,k_2)\right\rangle &=&2(Q+Q^{-1})\left| \Omega
(0,k_2)\right\rangle +\epsilon \left| \Omega (0,k_2-1)\right\rangle
+\epsilon \left| \Omega (0,k_2+1)\right\rangle  \nonumber \\
&&+\epsilon \left| \Omega (1,k_2)\right\rangle +(-\epsilon Q)^{N-2}\epsilon
\left| \Omega (k_2,N-1)\right\rangle  \label{cba35}
\end{eqnarray}
\begin{eqnarray}
{\cal H}\left| \Omega (k_1,N)\right\rangle &=&2(Q+Q^{-1})\left| \Omega
(k_1,N)\right\rangle +\epsilon \left| \Omega (k_1-1,N)\right\rangle
+\epsilon \left| \Omega (k_1+1,N)\right\rangle  \nonumber \\
&&+\epsilon \left| \Omega (k_1,N-1)\right\rangle +(-\epsilon
Q)^{-N+2}\epsilon \left| \Omega (1,k_1)\right\rangle  \label{cba36}
\end{eqnarray}
\begin{equation}
{\cal H}\left| \Omega (k,k+1\right\rangle =(Q+Q^{-1})\left| \Omega
(k,k+1\right\rangle +\epsilon \left| \Omega (k-1,k+1\right\rangle +\epsilon
\left| \Omega (k,k+2\right\rangle  \label{cba37}
\end{equation}
{\sf Substituting these results into the eigenvalue equation, we get the
following equations for wavefunctions corresponding to the separated
pseudoparticles. } 
\begin{eqnarray}
(E_{4\omega }-2Q-2Q^{-1})A(k_1,k_2) &=&\epsilon A(k_1-1,k_2)+\epsilon
A(k_1+1,k_2)  \nonumber \\
&&+\epsilon A(k_1,k_2-1)+\epsilon A(k_1,k_2+1)  \label{cba38}
\end{eqnarray}
{\sf i.e., for }$k_1\geq 1${\sf \ and }$k_1+3\leq k_2\leq N-1${\sf . The
boundary conditions read now } 
\begin{equation}
A(k_2,N+k_1)=(-\epsilon Q)^{N-2}A(k_1,k_2).  \label{cba39}
\end{equation}
{\sf The parametrization for the wavefunctions}

\begin{equation}
A(k_1,k_2)=A_{12}\xi _1^{k_1}\xi _2^{k_2}+A_{21}\xi _1^{k_2}\xi _2^{k_1}
\label{cba40}
\end{equation}
{\sf solves the equation (\ref{cba38}) provided that } 
\begin{equation}
E_{4s}=2(Q+Q^{-1})+\epsilon (\xi _1+\xi _1^{-1}+\xi _2+\xi _2^{-1})
\label{cba41}
\end{equation}
{\sf and the boundary conditions (\ref{cba39}) provided that } 
\begin{equation}
\xi _2^N=(-\epsilon Q)^{N-2}\frac{A_{21}}{A_{12}}\quad ,\quad \xi
_1^N=(-\epsilon Q)^{N-2}\frac{A_{12}}{A_{21}}\Rightarrow \xi ^N=(-\epsilon
Q)^{2(N-2)}  \label{cba42}
\end{equation}
{\sf where }$\xi =\xi _1\xi _2=e^{i(\theta _1+\theta _2)}${\sf , }$\theta
_1+\theta _2${\sf \ being the total momenta.}

{\sf Now we include the new states (\ref{cba34}) into the definition of }$%
\Psi _{4\omega }${\sf \ in order to extend (\ref{cba22}) to } 
\begin{equation}
\Psi _{4\omega }=\sum_{k_1<k_2}A(k_1,k_2)\left| \Omega (k_1,k_2\right\rangle
.  \label{cba43}
\end{equation}
{\sf Here we have used the same notation for separated and neighboring
states.}

{\sf Substituting (\ref{cba28}) and (\ref{cba37}) into the eigenvalue
equation, we get } 
\begin{equation}
(E_{4\omega }-Q-Q^{-1})A(k,k+1)=\epsilon A(k-1,k+1)+\epsilon A(k,k+2)
\label{cba44}
\end{equation}
{\sf which gives us the phase shift produced by the interchange of the two
pseudoparticles } 
\begin{equation}
\frac{A_{21}}{A_{12}}=-\frac{1+\xi +\epsilon (Q+Q^{-1})\xi _2}{1+\xi
+\epsilon (Q+Q^{-1})\xi _1}.  \label{cba45}
\end{equation}
{\sf We thus arrive to the Bethe ansatz equations which fix the values of }$%
\xi _1${\sf \ and }$\xi _2${\sf :}

\begin{eqnarray}
\xi _2^N &=&(-\epsilon Q)^{N-2}\left\{ -\frac{1+\xi +\epsilon (Q+Q^{-1})\xi
_2}{1+\xi +\epsilon (Q+Q^{-1})\xi _1}\right\} ,  \nonumber \\
\quad \xi _1^N\xi _2^N &=&(-\epsilon Q)^{2(N-2)}  \label{cba46}
\end{eqnarray}

\subsection{\sf General eigenstates}

{\sf Thus in the sector }$r=2\omega p${\sf , we expect that the }$p${\sf %
-pseudoparticle phase shift will be a sum of two-pseudoparticle phase shifts
and the energy is given by } 
\begin{equation}
E_{p(2s)}=\sum_{n=1}^p\left\{ Q+Q^{-1}+\epsilon (\xi _n+\xi _n^{-1})\right\}
\label{cba48}
\end{equation}
{\sf where } 
\[
\xi _a^N=(-\epsilon _sQ)^{N-2p+2}\prod_{b\neq a}^p\left\{ -\frac{1+\xi _a\xi
_b+\epsilon (Q+Q^{-1})\xi _a}{1+\xi _a\xi _b+\epsilon (Q+Q^{-1})\xi _b}%
\right\} ,\quad a=1,...,p 
\]
\begin{equation}
\left( \xi _1\xi _2\cdots \xi _p\right) ^N=(-\epsilon Q)^{p(N-2p+2)}
\label{cba49}
\end{equation}
{\sf The corresponding eigenstates are } 
\begin{equation}
\Psi _r(\xi _1,\xi _2,...\xi _p)=\sum_{1\leq k_1<...<k_p\leq
N-1}A(k_1,k_{2,}...,k_p)\left| \Omega (k_1,k_2,...,k_p)\right\rangle
\label{cba49a}
\end{equation}
{\sf where }$\left| \Omega (k_1,k_2,...,k_p)\right\rangle =\otimes
_{i=1}^p\left| \Omega (k_i)\right\rangle ${\sf \ and the wavefunctions
satisfy the following boundary conditions } 
\begin{equation}
A(k_1,k_{2,}...,k_p,N+k_1)=(-\epsilon Q)^{N-2p+2}A(k_1,k_{2,}...,k_p)
\label{cba49b}
\end{equation}

{\sf It is not all, in a sector }$r${\sf \ we may have }$p${\sf \
pseudoparticle and }$N_{\omega -1},N_{\omega -2},...,N_{-\omega +1}${\sf \
impurities of the type }$(\omega -1),(\omega -2),...,(-\omega +1)${\sf ,
respectively, such that } 
\begin{equation}
N_{\omega -1}+2N_{\omega -2}+\cdots +(2\omega -1)N_{-\omega +1}=r-2\omega p
\label{cba50}
\end{equation}
{\sf We called impurity a state }$\left| \alpha ,k\right\rangle ${\sf \
flanked by at least two states }$\left| \beta ,k\pm 1\right\rangle ${\sf \
such that }$\alpha +\beta \neq 0${\sf . Since }${\cal H}${\sf \ is a sum of
projectors on spin zero, these states are also annihilated by }${\cal U}_0$%
{\sf \ . Therefore the impurities play here the same role as in the periodic
case. It means that for a sector }$r${\sf \ with }$l${\sf \ impurities with
parameters }$\xi _1,...,\xi _l${\sf \ and }$p${\sf \ pseudoparticles with
parameters }$\xi _{l+1},...,\xi _{l+p}${\sf \ the energy is given by (\ref
{cba49}), and the Bethe equations do not depend on impurity type and are
given by } 
\begin{equation}
\xi _a^N\xi _1^2\xi _2^2\cdots \xi _l^2=(-\epsilon Q)^{N-2p+2}\prod_{b\neq
a=l+1}^{l+p}\left\{ -\frac{1+\xi _a\xi _b+\epsilon (Q+Q^{-1})\xi _a}{1+\xi
_a\xi _b+\epsilon (Q+Q^{-1})\xi _b}\right\}  \label{cba51}
\end{equation}
{\sf with }$a=l+1,l+2,...,l+p\quad ,\quad p\geq 1${\sf , and } 
\begin{equation}
\xi ^{2p}(\xi _{l+1}\cdots \xi _{l+p})^{N-2p}=(-\epsilon Q)^{p(N-2p+2)}
\label{cba52}
\end{equation}
{\sf where }$\xi =\xi _1\xi _2\cdots \xi _l\xi _{l+1}\cdots \xi _{l+p}${\sf .%
}

{\sf We have shown that these closed Temperley-Lieb quantum invariant spin
chains can be solved by the coordinate Bethe ansatz. A consequence of the
nonlocal terms }${\cal U}_0${\sf \ is the arising of boundary conditions
depending on the quantum group parameter }$q${\sf \ via the relation }$%
Q+Q^{-1}=${\sf Tr}$_{V_\Lambda }${\sf \ }$(q^{-2\rho }).${\sf \ It is also }$%
p${\sf -pseudoparticle dependent (which is equal to spin sector }$r${\sf \
for }$A_1${\sf , when }$s=1/2{\sf )}$.

{\sf For the algebra }$A_1$, {\sf and \ }$s=1/2${\sf \ , Q=q and }$U_k${\sf \
are 4x4 matrices giving a nearest-neighbour interaction }$U_k=q-\sigma
_k^{+}\sigma _{k+1}^{-}-\sigma _k^{-}\sigma _{k+1}^{+}-(q+q^{-1})/4(\sigma
_k^z\sigma _{k+1}^z+1)+(q+q^{-1})/4(\sigma _k^z-\sigma _{k+1}^z-2).$ {\sf %
This Hamiltonian was investigated in Ref.\cite{Pallua} and we summarize some
basic results regarding this case. Assuming q=}$\exp (i\varphi )${\sf \ some
interesting properties were found. For instance, the spin }$L${\sf \ of the
ground state becomes }$\varphi ${\sf \ - dependent. For any N (even), }$L$%
{\sf \ depends on the value of }$\varphi ${\sf \ according to :}

\begin{eqnarray*}
L &=&0 \mbox{\hspace{1.1cm} {\sf for} \hspace{1.38cm}}\frac{\pi}{2} \ < \ \varphi \ <  \ \pi \\
L &=&l \mbox{\hspace{1.16cm} {\sf for}\ \ }\frac{\pi}{2(l+1)} \ < \ \varphi \ < \ \frac{\pi}{2l} \\
L &=&\frac{N}{2} \mbox{\hspace{.86cm} {\sf for}\hspace{1.5cm}}0 \ < \ \varphi \ < \ \frac{\pi}{N} 
\end{eqnarray*}

{\sf The ground state is non-degenerate (up to the trivial SU}$_q${\sf \
degeneracy). At the edges of the intervals, }$\varphi ${\sf \ = }$\pi /2l$%
{\sf , additional degeneracies occur. These transitions to higher spins
resemble the incommensurate transition obtained in various other models.}

{\sf From the statistical mechanics point of view the Hamiltonian presents
critical behavior and it is conformal invariant. The central charge ( or
conformal anomaly ) is : } 
\[
c = 1 - \frac{6(\pi -\varphi)^2}{\pi \varphi} \, ,
\ \ \  \varphi \in [\pi/2,\pi ].  
\]
{\sf In particular, if we choose the rational form : } 
\[
\varphi = \frac{\pi m}{(m+1)} \, ,\mbox{\  \ \ }m=3,4...,
\]
{\sf then}
\[
c=1-\frac{6}{m(m+1)} \, ,
\]
{\sf which give us the conformal anomalies of the minimal unitary models.}

{\sf The connection between the Hamiltonian of the closed SU}$_q${\sf (2)
invariant chain and the unitary minimal series was explored in \cite{Pallua
1}. For a generic irrational }$\varphi ${\sf \ one can decompose the space of
states into the direct sum of irreducible representations of the quantum
group which are in one-to-one correspondence with the usual SU(2)
representations.}

{\sf We have constructed and diagonalized numerically the Hamiltonian for
small values of N and s=1/2, 1. We checked that, for a given N , s=1/2 and 1
have the same spectra up to degeneracies. \ }

\section{\sf \ Bethe Ansatz: Free boundary conditions}

{\sf It is for free boundary conditions that the Hamiltonian (\ref{int2})
naturally commutes with the quantum group }${\cal U}_q(X_n)${\sf . Since the
our linear combination (\ref{pba2}) left all models with the same status,
which concern to the coordinate Bethe ansatz, we expect that all procedure
developed for the coordinate Bethe ansatz with free boundary conditions in 
\cite{Alcaraz} for the case }$A_1${\sf \ (}$s=1/2${\sf ). can be used here.
To show this we recall the previous section, taking into account }${\cal U}%
_0=0${\sf , where almost all equations can be seized for the free boundary
conditions eigenvalue problem.}

\subsection{\sf One-pseudoparticle}

{\sf In this sector, the eigenstate is given by (\ref{cba9}):}

\begin{equation}
\Psi _{2\omega }(\xi )=\sum_{k=1}^{N-1}A(k)\left| \Omega (k)\right\rangle
\label{fba1}
\end{equation}
{\sf where }$\left| \Omega (k)\right\rangle ${\sf \ is again given by (\ref
{cba3}).}

{\sf The action of }${\cal H}${\sf \ on the states }$\left| \Omega
(k)\right\rangle ${\sf \ is given by (\ref{cba10}), which gives us the
following eigenvalue equations } 
\begin{equation}
(E_{2\omega }-Q-Q^{-1})A(k)=\epsilon A(k-1)+\epsilon A(k+1),\quad 2\leq
k\leq N-2  \label{fba2}
\end{equation}
{\sf At the boundaries, we get more two slightly different equations } 
\begin{eqnarray}
(E_{2\omega }-Q-Q^{-1})A(1)=\epsilon A(2)  \nonumber \\
(E_{2\omega }-Q-Q^{-1})A(N-1)=\epsilon A(N-2)  \label{fba3}
\end{eqnarray}
{\sf where }$\epsilon =1${\sf \ for }$B_n${\sf , }$D_n${\sf \ and }$A_1${\sf %
\ ( }$s${\sf \ integer) and }$\epsilon =-1${\sf \ for }$C_n${\sf \ and }$A_1$%
{\sf \ ( }$s${\sf \ semi-integer). We now try as a solution } 
\begin{equation}
A(k)={\rm A}(\theta )\xi ^k-{\rm A}(-\theta )\xi ^{-k}  \label{fba4}
\end{equation}
{\sf where }$\xi =e^{i\theta }${\sf , }$\theta ${\sf \ being the momenta.
Substituting this in equation (\ref{fba2}) we obtain the energy eigenvalue
associated with a free pseudoparticle with free boundary conditions } 
\begin{equation}
E_{2\omega }=Q+Q^{-1}+\epsilon \left( \xi +\xi ^{-1}\right)  \label{fba5}
\end{equation}

{\sf We want equations (\ref{fba2}) to be valid for }$k=1${\sf \ and }$k=N-1$%
{\sf \ also, where }$A(0)${\sf \ and }$A(N)${\sf \ are defined by (\ref{fba4}%
). Matching (\ref{fba2}) and (\ref{fba3}) we get the end conditions } 
\begin{equation}
A(0)=0\quad {\rm and}\quad A(N)=0  \label{fba6}
\end{equation}
{\sf implying that }$A(\theta )=A(-\theta )${\sf \ and }$\xi ^{2N}=1${\sf ,
respectively. }$A(\theta )${\sf \ it now determined ( up to a factor that is
invariant under }$\theta \longleftrightarrow -\theta ${\sf ), to be equal to 
}$\xi ^{-N}${\sf .}

\subsection{\sf One pseudoparticle and impurities}

{\sf Differently from the previous cases, due to the lack of periodicity,
the impurity positions are fixed. So, they have a different role in the
eigenvalue problem with free boundary conditions. For instance, let us
consider the case of one impurity of the type }$\omega -1${\sf , with
parameter }$\xi _1${\sf \ and one pseudoparticle with parameter }$\xi _2$%
{\sf . This eigenstate lies in the sector }$r=2\omega +1${\sf \ and we can
write } 
\begin{equation}
\Psi _{2\omega +1}(\xi _1,\xi _2)=\sum_{k_1<k_2}\left\{ A_1(k_1,k_2)\left|
\Omega _1(k_1,k_2)\right\rangle +A_2(k_1,k_2)\left| \Omega
_2(k_1,k_2)\right\rangle \right\}  \label{fba7}
\end{equation}
{\sf where }$\left| \Omega _i(k_1,k_2)\right\rangle ,\ i=1,2${\sf \ are
given by (\ref{pba10}).}

{\sf For this case we obtain the following eigenvalue equations } 
\begin{eqnarray}
(E_{2\omega +1}-Q-Q^{-1})A_1(k_1,k_2) &=&\epsilon A_1(k_1-1,k_2)+\epsilon
A_1(k_1+1,k_2)  \nonumber \\
(E_{2\omega +1}-Q-Q^{-1})A_2(k_1,k_2) &=&\epsilon A_2(k_1,k_2-1)+\epsilon
A_2(k_1,k_2+1)  \label{fba8a}
\end{eqnarray}
{\sf We also have two meeting conditions that arise because pseudoparticle
and impurity may be neighbors (see (\ref{pba21})) } 
\begin{equation}
A_1(k,k)=A_2(k,k+2)\ ,\ A_2(k+1,k+2)=A_1(k,k+1)  \label{fba9a}
\end{equation}
{\sf in addition to the two conditions to be satisfied at the free ends } 
\begin{equation}
A_1(k_1,N)=0\quad ,\quad A_2(0,k_2)=0  \label{fba10a}
\end{equation}
{\sf Now we try the following ansatz for the wavefunctions } 
\begin{eqnarray}
A_1(k_1,k_2) &=&{\rm A}_1(\theta _1,\theta _2)\xi _1^{k_1}\xi _2^{k_2}-{\rm A%
}_1(\theta _1,-\theta _2)\xi _1^{k_1}\xi _2^{-k_2}  \nonumber \\
A_2(k_1,k_2) &=&{\rm A}_2(\theta _1,\theta _2)\xi _2^{k_1}\xi _1^{k_2}-{\rm A%
}_2(\theta _1,-\theta _2)\xi _2^{-k_1}\xi _1^{k_2}  \label{fba11a}
\end{eqnarray}
{\sf From (\ref{fba8a}) we get the energy eigenvalue } 
\begin{equation}
E_{2\omega +1}=Q+Q^{-1}+\epsilon \left( \xi _2+\xi _2^{-1}\right)
\label{fba12a}
\end{equation}
{\sf and from (\ref{fba9a}) and (\ref{fba10a}) the following relations
between the coefficients }$A_i$%
\begin{eqnarray}
{\rm A}_1(\theta _1,\theta _2)\xi _2^N &=&{\rm A}_1(\theta _1,-\theta _2)\xi
_2^{-N}\quad ,\quad {\rm A}_2(\theta _1,\theta _2)={\rm A}_2(\theta
_1,-\theta _2)  \nonumber \\
{\rm A}_1(\theta _1,\theta _2) &=&{\rm A}_2(\theta _1,\theta _2)\xi
_1^2\quad ,\quad {\rm A}_1(\theta _1,-\theta _2)={\rm A}_2(\theta _1,-\theta
_2)\xi _1^2  \label{fba13a}
\end{eqnarray}
{\sf from this we get } 
\begin{equation}
\xi _2^{2N}=1  \label{fba14a}
\end{equation}
{\sf as the Bethe equation of (\ref{fba12a}). The coefficients }$A_i${\sf \
are determined up to a factor that is invariant under }$\theta
_2\longleftrightarrow -\theta _2${\sf \ as: } 
\begin{equation}
{\rm A}_1(\theta _1,\theta _2)=\xi _1^2\xi _2^{-N}\qquad {\rm and}\qquad 
{\rm A}_2(\theta _1,\theta _2)=\xi _2^{-N}.  \label{fba15a}
\end{equation}
{\sf In general, for the eigenstate with }$l${\sf \ impurities with
parameters }$\xi _1,...,\xi _l${\sf \ and one pseudoparticle with parameter }%
$\xi _{l+1}${\sf , which lies in a sector }$r${\sf , we can write } 
\begin{equation}
\Psi _r(\xi _1,...,\xi _{l+1})=\sum_{j=1}^{l+1}\left\{ \sum_{1\leq
k_1<...<k_{l+1}\leq N-1}A_j(k_1,...,k_{l+1})\left| \Omega
_j(k_1,...,k_{l+1})\right\rangle \right\}  \label{fba16a}
\end{equation}
{\sf The corresponding eigenvalue is given by (\ref{fba5}) , with }$\xi =\xi
_{l+1}${\sf , and the ansatz for the coefficients of the wavefunctions
becomes } 
\begin{equation}
{\rm A}_j(\theta _1,...,\theta _{l+1})=\left( \prod_{i=1}^{l+1-j}\xi
_i^2\right) \xi _{l+1}^{-N}  \label{fba17a}
\end{equation}
{\sf Here we notice that the index }$j${\sf \ in the wavefunctions }$%
A_j(k_1,...,k_{l+1})${\sf \ means that the pseudoparticle is at the position 
}$k_{l+2-j}${\sf .}

\subsection{\sf Two-pseudoparticles}

{\sf For the sector }$r=4\omega ${\sf , beside eigenstates with impurities,
we have an eigenstate with two pseudoparticles. We obtain the following
eigenvalue equations } 
\begin{eqnarray}
(E_{4\omega }-2Q-2Q^{-1})A(k_1,k_2)=\epsilon A(k_1-1,k_2)+\epsilon
A(k_1+1,k_2)  \nonumber \\
+\epsilon A(k_1,k_2-1)+\epsilon A(k_1,k_2+1)  \label{fba8}
\end{eqnarray}
{\sf We have again two conditions to be satisfied at the ends of the chain } 
\begin{equation}
A(0,k_2)=0\quad {\rm and}\quad A(k_1,N)=0  \label{fba9}
\end{equation}
{\sf In addition to this we have a meeting condition } 
\begin{equation}
\epsilon A(k,k)+\epsilon A(k+1,k+1)+(Q+Q^{-1})A(k,k+1)=0  \label{fba10}
\end{equation}

{\sf Now we try the ansatz } 
\begin{equation}
\begin{array}{lll}
A(k_1,k_2) & = & {\rm A}(\theta _1,\theta _2)\xi _1^{k_1}\xi _2^{k_2}-{\rm A}%
(\theta _2,\theta _1)\xi _1^{k_2}\xi _2^{k_1} \\ 
& - & {\rm A}(-\theta _1,\theta _2)\xi _1^{-k_1}\xi _2^{k_2}+{\rm A}(-\theta
_2,\theta _1)\xi _1^{-k_2}\xi _2^{k_1} \\ 
& - & {\rm A}(\theta _1,-\theta _2)\xi _1^{k_1}\xi _2^{-k_2}+{\rm A}(\theta
_2,-\theta _1)\xi _1^{k_2}\xi _2^{-k_1} \\ 
& + & {\rm A}(-\theta _1,-\theta _2)\xi _1^{-k_1}\xi _2^{-k_2}-{\rm A}%
(-\theta _2,-\theta _1)\xi _1^{-k_2}\xi _2^{-k_1}
\end{array}
\label{fba11}
\end{equation}
{\sf Here we observe the permutations and negations of }$\theta _1${\sf \
and }$\theta _2${\sf . Substituting this ansatz in (\ref{fba8}) we obtain
the energy eigenvalue for the sector with two pseudoparticles } 
\begin{equation}
E_{4\omega }=2Q+2Q^{-1}+\epsilon \left( \xi _1+\xi _1^{-1}+\xi _2+\xi
_2^{-1}\right)  \label{fba12}
\end{equation}
{\sf The ansatz (\ref{fba11}) satisfy equations (\ref{fba9}) provided that } 
\begin{eqnarray}
{\rm A}(\theta _1,\theta _2)={\rm A}(-\theta _1,\theta _2)\quad ,\quad {\rm A%
}(\theta _2,\theta _1)={\rm A}(-\theta _2,\theta _1)  \nonumber \\
{\rm A}(\theta _1,-\theta _2)={\rm A}(-\theta _1,-\theta _2)\ ,\quad {\rm A}%
(\theta _2,-\theta _1)={\rm A}(-\theta _2,-\theta _1)  \label{fba13}
\end{eqnarray}
{\sf and } 
\begin{equation}
\xi _2^{2N}=\frac{{\rm A}(\theta _1,-\theta _2)}{{\rm A}(\theta _1,\theta _2)%
}=\frac{{\rm A}(-\theta _1,-\theta _2)}{{\rm A}(-\theta _1,\theta _2)},\quad
\xi _1^{2N}=\frac{{\rm A}(\theta _2,-\theta _1)}{{\rm A}(\theta _2,\theta _1)%
}=\frac{{\rm A}(-\theta _2,-\theta _1)}{{\rm A}(-\theta _2,\theta _1)}
\label{fba14}
\end{equation}
{\sf Moreover, the meeting conditions are satisfied provided that } 
\begin{eqnarray}
\frac{{\rm A}(-\theta _1,-\theta _2)}{{\rm A}(-\theta _2,-\theta _1)}=\frac{%
{\rm A}(\theta _2,\theta _1)}{{\rm A}(\theta _1,\theta _2)}=\frac{1+\xi
_1\xi _2+\epsilon \left( Q+Q^{-1}\right) \xi _2}{1+\xi _1\xi _2+\epsilon
\left( Q+Q^{-1}\right) \xi _1}  \nonumber \\
\frac{{\rm A}(\theta _1,-\theta _2)}{{\rm A}(-\theta _2,\theta _1)}=\frac{%
{\rm A}(\theta _2,-\theta _1)}{{\rm A}(-\theta _1,\theta _2)}=\frac{1+\xi
_1^{-1}\xi _2+\epsilon \left( Q+Q^{-1}\right) \xi _2}{1+\xi _1^{-1}\xi
_2+\epsilon \left( Q+Q^{-1}\right) \xi _1^{-1}}  \label{fba15}
\end{eqnarray}
{\sf Matching these conditions we get } 
\begin{equation}
\xi _1^{2N}=\frac{B(-\theta _1,\theta _2)}{B(\theta _1,\theta _2)}\quad
,\quad \xi _2^{2N}=\frac{B(-\theta _2,\theta _1)}{B(\theta _2,\theta _1)}
\label{fba16}
\end{equation}
{\sf and } 
\begin{equation}
{\rm A}(\theta _1,\theta _2)=\xi _1^{-N}\xi _2^{-N}B(-\theta _1,\theta
_2)\xi _2^{-1}.  \label{fba17}
\end{equation}
{\sf Here we have used the usual free boundary notations } 
\begin{equation}
B(\theta _a,\theta _b)=s(\theta _a,\theta _b)\ s(\theta _b,-\theta _a)
\label{fba18}
\end{equation}
{\sf where } 
\begin{equation}
s(\theta _a,\theta _b)=1+\xi _a\xi _b+\epsilon \left( Q+Q^{-1}\right) \xi _b.
\label{fba19}
\end{equation}

{\sf Now let us consider the eigenstates with two pseudoparticle and
impurities. The energy eigenvalue is the same of the two pseudoparticles
pure state. The parameters associated with impurities are embraced in the
definition of the coefficients of the wavefunctions. For instance, when we
have an eigenstate of two pseudoparticles with parameters }$\xi _2${\sf \
and }$\xi _3${\sf \ and one impurity of parameter }$\xi _1${\sf , the energy
is given by (\ref{fba12}) and the Bethe equations by (\ref{fba16}), with }$%
\xi _1\rightarrow \xi _3${\sf \ and }$\theta _1\rightarrow \theta _3${\sf .
But now the wavefunctions are different } 
\begin{eqnarray}
{\rm A}_1(\theta _1,\theta _2,\theta _3)=\left\{ \xi _1^4\right\} \xi
_2^{-N}\xi _3^{-N}B(-\theta _2,\theta _3)\xi _3^{-1}  \nonumber \\
{\rm A}_2(\theta _1,\theta _2,\theta _3)=\left\{ \xi _1^2\right\} \xi
_2^{-N}\xi _3^{-N}B(-\theta _2,\theta _3)\xi _3^{-1}  \nonumber \\
{\rm A}_3(\theta _1,\theta _2,\theta _3)=\xi _2^{-N}\xi _3^{-N}B(-\theta
_2,\theta _3)\xi _3^{-1}  \label{fba20}
\end{eqnarray}
{\sf where }$B(-\theta _2,\theta _3)${\sf \ is given by (\ref{fba18})}

\subsection{\sf General eigenstates}

{\sf The generalization follows as in the previous cases. In a sector }$r$%
{\sf \ with}

$p${\sf \ pseudoparticles, we get } 
\begin{equation}
E_r=\sum_{n=1}^p\left[ Q+Q^{-1}+\epsilon \left( \xi _n+\xi _n^{-1}\right)
\right]  \label{fba21}
\end{equation}
{\sf and the Bethe equations } 
\begin{equation}
\xi _a^{2N}=\prod_{b\neq a=l+1}^{l+p}\frac{B(-\theta _a,\theta _b)}{B(\theta
_a,\theta _b)}\ ,\qquad a=1,2,...,p  \label{fba22}
\end{equation}
{\sf The corresponding eigenfunction can be written as } 
\begin{equation}
\Psi _r(\xi _1,...,\xi _p)=\sum_{k_1<\cdots
<k_{l+p}}A(k_1,k_2,...,k_{l+p})\left| \Omega (k_1,k_2,...,k_p)\right\rangle
\label{fba23}
\end{equation}
{\sf with } 
\begin{equation}
A(k_1,k_2,...,k_p)=\sum_P\varepsilon _P\ {\rm A}(\theta _1,\theta
_2,...,\theta _p)\ \xi _1^{k_1}\xi _2^{k_2}\cdots \xi _p^{k_p}  \label{fba24}
\end{equation}
{\sf where the sum extends over all permutations and negations of }$\theta
_1,...,\theta _p${\sf \ and }$\varepsilon _P${\sf \ changes sign at each
such interchange. The coefficients in the wavefunctions are given by } 
\begin{equation}
{\rm A}(\theta _1,\theta _2,...,\theta _p)=\prod_{j=1}^p\xi
_j^{-N}\prod_{l+1\leq j<i\leq l+p}B(-\theta _j,\theta _i)\xi _j^{-1}
\label{fba25}
\end{equation}
{\sf where }$B(-\theta _j,\theta _i)${\sf \ are defined in (\ref{fba18}).}

{\sf For a sector }$r${\sf \ with }$l${\sf \ impurities with parameters }$%
\xi _1,...,\xi _l${\sf \ and }$p${\sf \ pseudoparticles with parameters }$%
\xi _{l+1},...,\xi _{l+p}${\sf \ the energy is given by (\ref{fba21}) and
the Bethe equations by (\ref{fba22}). Only the coefficients of the wave
functions are modified } 
\begin{equation}
{\rm A}_j(\theta _1,\theta _2,...,\theta _{l+p})=A_j(\xi _1\xi _2\cdots \xi
_l){\rm A}(\theta _{l+1},\theta _2,...,\theta _{l+p}).  \label{fba26}
\end{equation}
{\sf The functions }$A_j(\xi _1\xi _2\cdots \xi _l)=\xi _1^{a_1}\xi
_2^{a_2}\cdots \xi _l^{a_l}${\sf \ where the index }$j${\sf \ characterizes
the possible configurations of }$l${\sf \ impurities relative to the }$p$%
{\sf \ pseudoparticles. Here }$a_i${\sf \ are numbers which depend on the
position of corresponding impurity relative to the pseudoparticles.}

{\sf Here we observe again the valid of these results for all Temperley-Lieb
spin chain Hamiltonians defined as projector of spin zero on the
representations of the quantum groups }${\cal U}_q(X_n)${\sf , characterized
by the values of }$Q+Q^{-1}={\rm Tr}_{V_\Lambda }(q^{-2\rho }).$

\section{\sf Conclusion}

{\sf We have applied in a systematic way the coordinate Bethe ansatz to find
the spectra of a series of ''spin '' Hamiltonians arising as representations
of the Temperley--Lieb algebra. We consider several boundary conditions in
order to include all previously known cases.}

{\sf Due to }${\cal U}_k${\sf \ be a projector of spin zero, there is a
linear combination of eigenstates of }${\bf S}_T^Z=\sum_k{\bf S}_k^Z${\sf \
where }${\bf S}_k^Z={\rm diag}(\omega ,...,-\omega ),\ \omega =\max (J)${\sf %
, which beside simplify the calculus permits us a unified treatment for all
models. We find that for a given set of boundary conditions, all models have
equivalent spectra, i.e. they differ at most in their degeneracies.
Moreover, for the closed cases, the spectra of the lower-dimensional
representations are entirely contained in the higher-dimensional ones (see
Eq.(\ref{pba23})). }

{\sf Here we notice that this spectrum equivalence is, of course, a
consequence of the TL algebra. Nevertheless there is in the literature a
large class of Hamiltonians which are not derived from representations of
the TL algebra which share the same property. The authors of reference \cite
{HS} developed a technique for construction of spin chain Hamiltonians which
affine quantum group symmetry whose spectra coincides with the spectra of
spin chain Hamiltonians which have non-affine quantum group symmetry.}

{\sf The energy eigenvalues are given by } 
\begin{eqnarray}
E &=&\sum_{n=1}^p\left( Q+Q^{-1}+2\epsilon \cos \theta _n\right)  \nonumber
\\
Q+Q^{-1} &=&{\rm Tr}_{V_\Lambda }(q^{-2\rho })  \label{c1}
\end{eqnarray}
{\sf where }$\rho ${\sf \ is half the sum of the positive roots of }${\cal U}%
_q(X_n),\quad X_n=A_1,B_n,C_n\ ${\sf and}$\ D_n${\sf . }$\theta _n${\sf \
are solutions of the Bethe ansatz equations (\ref{pba60}), (\ref{cba51}) and
(\ref{fba22}).}

{\sf The Hamiltonians for the cases }$X_n=B_n,C_n\ ${\sf and}$\ D_n${\sf \
appear to be new although due to the Temperley--Lieb equivalence \cite{B},
they are expected to possess the same thermodynamic properties as the }$%
A_1(s=1/2)${\sf \ case, i.e., the spin-}$1/2${\sf \ XXZ chain with
appropriate coupling.}

{\sf There are several issues left for future work. In particular, one would
like clarify from an algebraic point of view, the equality of the spectra,
for instance, of the biquadratic model (}$A_1,s=1${\sf ) and XXZ model (}$%
A_1,s=1/2${\sf \ ) for free boundary conditions and the inclusion of the XXZ
spectrum in the one of the biquadratic Hamiltonian for periodic boundary
conditions. Furthermore, the completeness and complete characterization as
highest weight states of the Bethe ansatz eigenstates here presented are not
considered.}

{\sf Using our solutions, one can derive partition functions in the
finite--size scaling limit and find the operator content of the systems
constructed from these quantum chains.}

{\sf Finally, we remark here that although the Hamiltonian (\ref{cbah}) is a
global operator, it manifests the property of essential locality \cite{Links}%
. From the physical point of view, this type of models exhibit behavior
similar to closed chains with twisted boundary conditions, however now the
boundary conditions become sector dependent.}

{\sf Acknowledgments: We would like to thank A. Lima--Santos for helpful
comments.This work was supported in part by CAPES-PICD, Conselho Nacional
de Desenvolvimento Cient\'\i fico e Tecnol\'ogico (CNPq-Brazil) (RCTG),
and by Funda\c c\~ao de Amparo \`a Pesquisa do Estado de S\~ao Paulo
(FAPESP-Brazil) (ALM).}

\end{document}